\documentclass[twocolumn]{aastex6}
\usepackage{times,color,natbib,url,amsmath}
\usepackage{graphicx,float,psfrag}
\usepackage{tabularx}
\usepackage{latexsym,amsmath,amssymb}
\usepackage{natbib}
\usepackage{verbatim}
\usepackage{multirow}
\usepackage{color}
\usepackage{enumitem}

\bibpunct{(}{)}{;}{a}{}{,}

\newcommand {\asca} {{\it ASCA}}

\newcommand {\xmm} {\textsl{XMM-Newton}}

\newcommand {\nustar} {\textsl{NuSTAR}}

\def \rsun {\ifmmode$R$_{\odot}\else R$_{\odot}$}

\def \hcm {\hbox {\ifmmode $ atoms cm$^{-2}\else atoms cm$^{-2}$\fi}}

\def\approxgt{\mathrel{\hbox{\rlap{\lower.55ex \hbox {$\sim$}}
        \kern-.3em \raise.4ex \hbox{$>$}}}}
\def\approxlt{\mathrel{\hbox{\rlap{\lower.55ex \hbox {$\sim$}}
        \kern-.3em \raise.4ex \hbox{$<$}}}}

\newcommand {\msun} {{{\rm M$_{\odot}$}}}
\newcommand {\degree} {$^{\circ}$}
\def \arcmin {\hbox{$^\prime$}}
\def \arcsec {\hbox{$^{\prime\prime}$}}

\def \srco {NGC~3998}
\def \srct {NGC~4579}

\begin{document}

\title{\nustar\ Hard X-ray View of Low-luminosity Active Galactic
  Nuclei:\\High-energy Cutoff and Truncated Thin Disk}

%
% Authors
%% Authors with the same affiliation can be grouped in a single
%% \author and \affil call.
\author{George~Younes\altaffilmark{1,2}}
\author{Andrew~Ptak\altaffilmark{3}}
\author{Luis~C.~Ho\altaffilmark{4,5}}
\author{Fu-Guo~Xie\altaffilmark{6}}
\author{Yuichi~Terasima\altaffilmark{7}}
\author{Feng~Yuan\altaffilmark{6}}
\author{Daniela Huppenkothen\altaffilmark{8}}
\author{Mihoko Yukita\altaffilmark{3,9}}

\altaffiltext{1}{Department of Physics, The George Washington University, Washington, DC 20052, USA, gyounes@gwu.edu}
\altaffiltext{2}{Astronomy, Physics and Statistics Institute of Sciences (APSIS), The George Washington University, Washington, DC 20052, USA}
\altaffiltext{3}{NASA Goddard Space Flight Center, Code 662, Greenbelt, MD 20771, USA}
\altaffiltext{4}{Kavli Institute for Astronomy and Astrophysics, Peking University, Beijing 100871, Peopleʼs Republic of China}
\altaffiltext{5}{Department of Astronomy, School of Physics, Peking University, Beijing 100871, Peopleʼs Republic of China}
\altaffiltext{6}{Key Laboratory for Research in Galaxies and Cosmology, Shanghai Astronomical Observatory, Chinese Academy of Sciences, 80 Nandan Road, Shanghai 200030, China}
\altaffiltext{7}{Department of Physics, Ehime University, Bunkyo-cho, Matsuyama, Ehime 790-8577, Japan}
\altaffiltext{8}{Department of Astronomy, University of Washington, 3910 15th Ave NE, Seattle, WA 98195}
\altaffiltext{9}{Johns Hopkins University, Homewood Campus, Baltimore, MD 21218, USA}

\begin{abstract}

We report the analysis of simultaneous \xmm+\nustar\ observations
of two low-luminosity Active Galactic Nuclei (LLAGN), \srco\ and
\srct. We do not detect any significant variability in either source
over the $\sim3$-day length of the \nustar\ observations. The
broad-band 0.5-60~k1eV spectrum of \srco\ is best fit with a cutoff
power-law, while the one for \srct\ is best fit with a combination of
a hot thermal plasma model, a power-law, and a blend of Gaussians to
fit an Fe complex observed between 6 and 7~keV. Our main spectral
results are the following: (1) neither source shows any reflection
hump with a $3\sigma$ reflection fraction upper-limits $R<0.3$ and
$R<0.18$ for \srco\ and \srct, respectively; (2) the 6-7~keV line
complex in \srct\ could either be fit with a narrow Fe~K line at
6.4~keV and a moderately broad Fe~XXV line, or 3 relatively narrow
lines, which includes contribution from Fe~XXVI; (3) \srct\ flux is
60\% brighter than previously detected with \xmm, accompanied by a
hardening in the spectrum; (4) we measure a cutoff energy $E_{\rm
  cut}=107_{-18}^{+27}$~keV in \srco, which represents the lowest and
best constrained high-energy cutoff ever measured for an LLAGN; (5)
\srco\ spectrum is consistent with a Comptonization model with
either a sphere ($\tau\approx3\pm1$) or slab ($\tau\approx1.2\pm0.6$)
geometry, corresponding to plasma temperatures between 20 and
150~keV. We discuss these results in the context of hard X-ray
emission from bright AGN, other LLAGN, and hot accretion flow models.

\end{abstract}

\section{Introduction}
\label{Intro}

%-----------
% Table  1
%-----------
\begin{table*}[]
\caption{\nustar+\xmm\ observations of \srco\ and \srct}
\label{logObs}
\newcommand\T{\rule{0pt}{2.6ex}}
\newcommand\B{\rule[-1.2ex]{0pt}{0pt}}
\begin{center}{
\resizebox{0.98\textwidth}{!}{
\hspace*{-1.5cm}
\begin{tabular}{l c c c c c c c}
\hline
\hline
Name & Optical Classification & Distance & $\log M_{\rm BH}$ & Telescope & Observation ID & Date & GTI$^a$ \\
  \T\B &                                    &  Mpc       & \msun\               &                  &                         &          &  ks \\
\hline
\srco\       \B  & LINER~1.9          & 14.1       & 9.3 & \nustar\ &  60201050002   &  2016/10/25 & 104/104 \\
                 \B  &                           &               &        & \xmm\   & 0790840101     &  2016/10/26 & 3.8/11.4/11.2 \\
\srct\ \B &LINER~1.9/Seyfert~1.9& 16.4       & 8.1  & \nustar\ & 60201051002   &  2016/12/06 & 118/117 \\
\B                   &                           &               &         & \xmm\   & 0790840201     &  2016/12/06 & 17.2/21.3/21.2 \\
\hline
\end{tabular}}}
\end{center}
\begin{list}{}{}
\item[{\bf Notes.}] Optical classification is taken from
  \citet{ho97apjs:broadHal}. Distances are from
  \citet{tonry01apj:dist} and \citet{Springob07ApJS:dist} for \srco\
  and \srct, respectively. Black-hole masses are calculated using
  velocity dispersions from \citet{ho09apjs:veldisp} and the $M-\sigma$
  relation of \citet{kormendy13ARAA}. $^a$ Good time intervals
  represent the live time of FPMA/FPMB, and the cleaned exposure for
  pn/mos1/mos2.
\end{list}
\end{table*}
%-----------
% Table  1
%-----------

Most galaxies in the nearby universe host a low luminosity active
galactic nucleus (LLAGN), with bolometric luminosities ranging from
$10^{38}$ to $10^{43}$~erg~s$^{-1}$ \citep{flohic06ApJ:liner,
    zhang09ApJ:liner,gonzalezmartin09aa}. Assuming that the emission
is due to accretion onto a supermassive black-hole with masses $M_{\rm
  BH}\approx10^{7}$-$10^{9}$~$M_{\odot}$, these bolometric
luminosities translate into Eddington ratios in the range
$10^{-7}$ to $10^{-3}$; at least an order of magnitude
smaller than the ratio for luminous active galactic nuclei (AGN). The
dimness of these LLAGN is due to an underfed supermassive black-hole,
and since these sources represent the bulk of active galaxies
\citep{ho97ApJ:487}, understanding the effect of low accretion rates
on the geometry and dynamics of the central engine of galaxies is
evidently imperative \citep[see, e.g.,][for reviews]{ho08aa:review,
  ho09apj:riaf}.

Multiwavelength observations of LLAGN have shown a particularly
different spectral energy distribution compared to luminous AGN. For
instance, unlike most bright AGN in the local universe, i.e., Seyfert
galaxies, LLAGN appear as radio-loud sources, with radio to X-ray flux
ratio comparable to radio-loud AGN. They also lack a UV bump, the
ubiquitous feature in almost all AGN \citep[e.g.,][]{ho99sed,
  4278nagar05aap,eracleous10ApJS,younes12AA}. Specifically in soft
X-rays, LLAGN do not show the strong, intra-day variability shared by
their more luminous counterparts \citep[e.g., ][]{ptak98apj:variance,
  pianmnras10,younes10aa:ngc4278,younes11AA:liner1sXray,
  gonzalez12AA:var,hernandez2014AA}, while their soft X-ray spectra
appear mostly featureless, especially lacking the strong reflection
features, e.g., a broad Fe~K$\alpha$ line \citep{ptak04apj:ngc3998,
  gonzalezmartin09aa,younes11AA:liner1sXray}. Furthermore, LLAGN do
not seem to follow some of the correlations established for typical
AGN; the positive $\Gamma-$Eddington-ratio correlation seen in
luminous AGN \citep{sobolewska09mnras:gamvsedd} does not extend down
to the limits of LLAGN, instead LLAGN show an opposite,
anticorrelation between the 2 parameters \citep[][]{
  gu09mnras:gamVSeddllagn,younes11AA:liner1sXray,yang15MNRAS,
  she18ApJ:llagn}. This overwhelming, yet incomplete, list of evidence
point towards a drastically altered central engine in LLAGN compared
to luminous AGN.

These unique properties of LLAGN cannot be understood in the context
of the radiatively efficient, geometrically thin accretion disks that are
thought to power luminous AGN \citep{shakura73aa}. For the low
accretion rates that govern LLAGN, the density in the accretion disk
becomes too low for radiative cooling to be effective. Hence, the
trapped heat will expand the inner parts of the accretion disk into a
pressure-supported, radiatively-inefficient hot accretion flow
\citep[see, e.g., ][for reviews]{narayan08:riafreview,yuan14ARAA}. The
most famous examples of such hot accretion solutions are the advection
dominated accretion flow \citep[e.g., ][]{narayan94ApJ:adaf}, the
adiabatic inflow-outflow solution \citep[e.g.,][]{
  blandford99MNRAS:jet}, and the convection dominated accretion flow
\citep{narayan00ApJ}. These models have been successfully fit to the
SED of a number of LLAGN \citep[e.g.,][]{dimatteo01ApJ:jetliner,
  ptak04apj:ngc3998,xu09RAA:adaf,nemmen14MNRAS}, including Sgr~A$^*$
\citep[e.g.,][]{yuan04ApJ,wang13Sci:sgrA}.

Hard X-ray observations of AGN are of paramount importance. The X-ray
emission of AGN emanates from a corona, where optical and UV photons
from the disk Compton up-scatter into the X-ray band \citep{
  haardt93ApJ:corona}. The temperature in the corona reveals itself
through a break in the hard X-ray energies. Moreover, a reflection
feature is expected in the hard band in the form of a ``Compton hump''
at energies of $\sim$30~keV. While the brightest AGN have been studied
in the past at those energies \citep[e.g., ][]{montavini16MNRAS},
\nustar, the first focusing X-ray telescope at hard X-rays, have
revolutionized the field, allowing us to study with unprecedented
details those signatures, not only in moderately bright AGN, but for
the first time in LLAGN as well. Indeed, at least two LLAGN have been
observed with \nustar\ so far, NGC~7213 and M~81 \citep{
  ursini15MNRAS:7213,young18MNRAS:m81}. Neither source showed any
hint of a relativistic disk reflection component in the hard X-ray
band, pointing towards a truncated inner accretion disk, possibly
filled with a hot accretion flow instead.

\begin{figure*}[t!]
\begin{center}
\includegraphics[angle=0,width=0.9\textwidth]{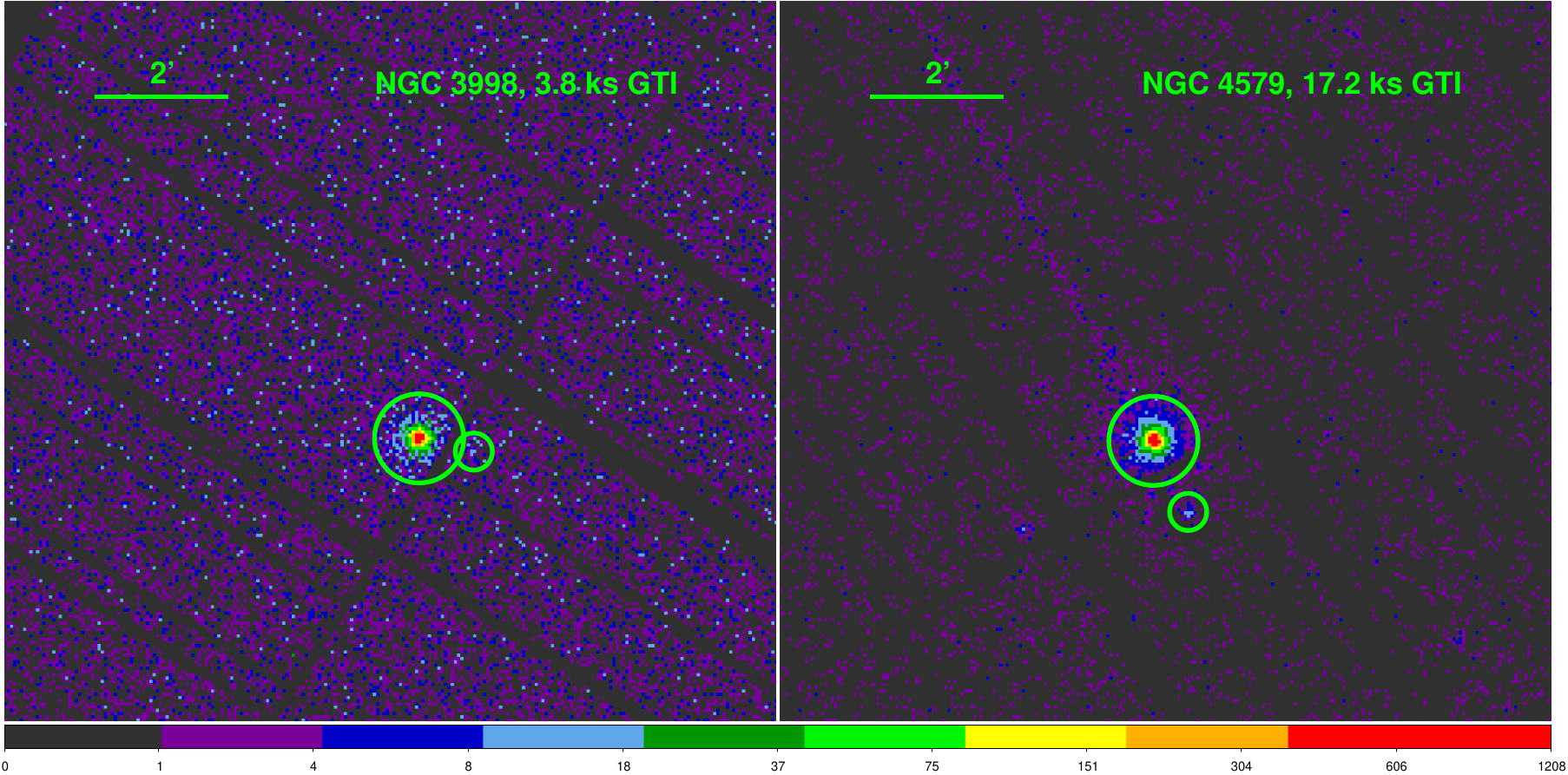}
\caption{\xmm\ EPIC-pn image of \srco\ ({\sl left panel}) and \srct\
  ({\sl right panel}) in the 2-10 keV energy range. The large green
  circles, with radii of 60\arcsec, are centered at the galactic
  nuclei. The other circular regions are off-nuclear X-ray point
  sources. The X-ray emission of these sources is $\lesssim1\%$ of the
  LLAGN. The color bar is in units of counts.}
\label{skyIm}
\end{center}
\hspace*{1.0cm}
\end{figure*}

Here, we report two deep, simultaneous \nustar+\xmm\ observations of
the LLAGN \srco\ and \srct. Both galaxies are optically classified as
type~1.9, showing broad H$\alpha$ emission lines
\citep[Table~\ref{logObs},][]{ho97apjs:broadHal}. They have been
previously studied in soft X-rays with \xmm\ \citep{
  ptak04apj:ngc3998,dewangan04ApJ}. Their radio to X-ray SEDs have
been successfully fit with an ADAF and/or a jet model \citep{
  ptak04apj:ngc3998,quataert99apj:m81ngc4579,xu09RAA:adaf}. \srco\ and
\srct\ are among the lowest Eddington ratio AGN to be observed at hard
X-rays $\gtrsim10$~keV, with $L_{\rm bol}/L_{\rm Edd}$ of
$1.0\times10^{-5}$ and $1.0\times10^{-4}$ for \srco\ and \srct,
respectively \citep{younes12AA,nemmen14MNRAS}. Section~\ref{obs}
details the observation and data analysis procedures. We present our
timing and spectral results in section~\ref{res}, which are discussed
in details in section~\ref{discuss}. Finally, a summary of our findings
is presented in section~\ref{conc}.

\begin{figure*}[t]
\begin{center}
\includegraphics[angle=0,width=0.49\textwidth]{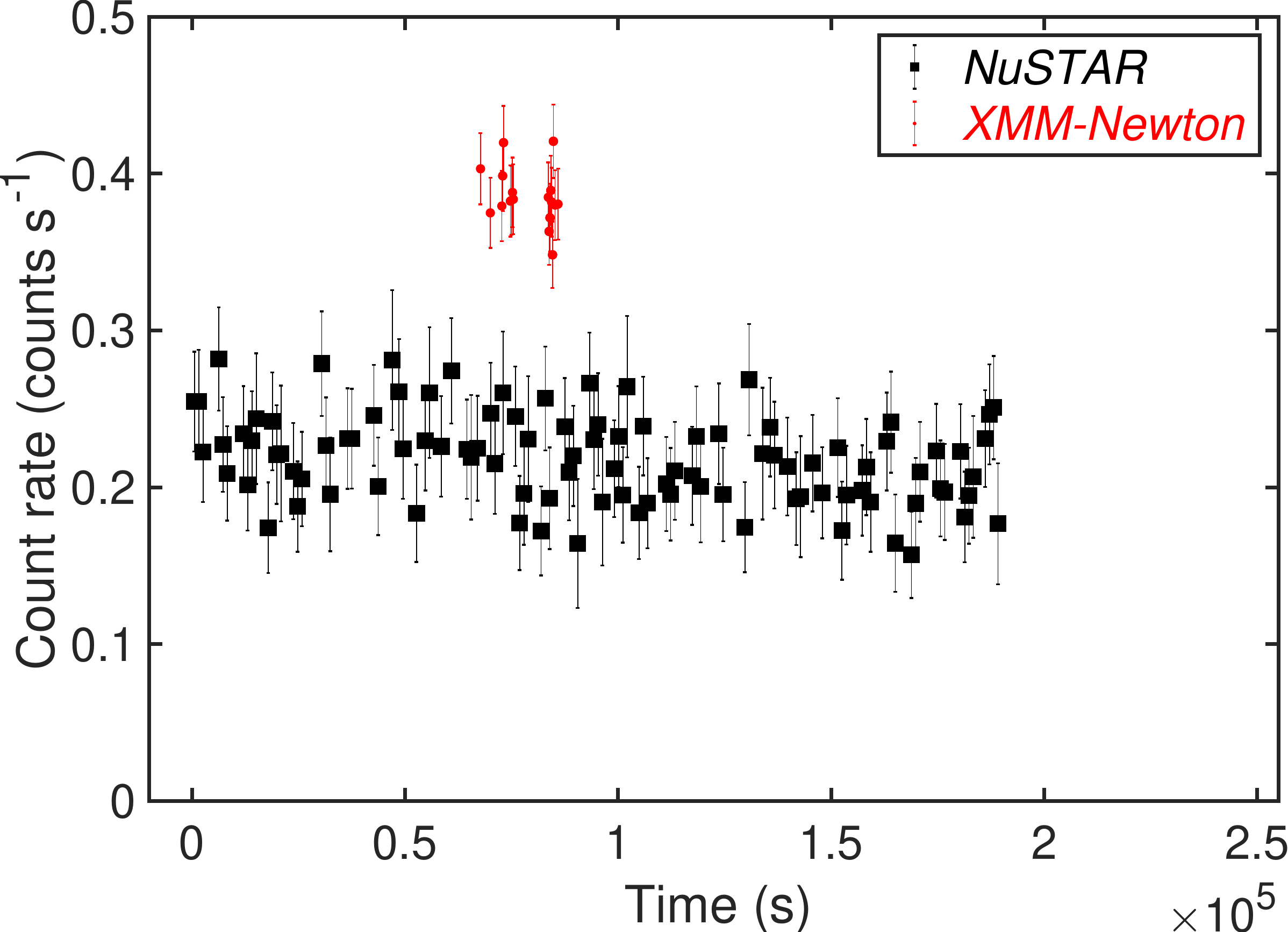}
\includegraphics[angle=0,width=0.49\textwidth]{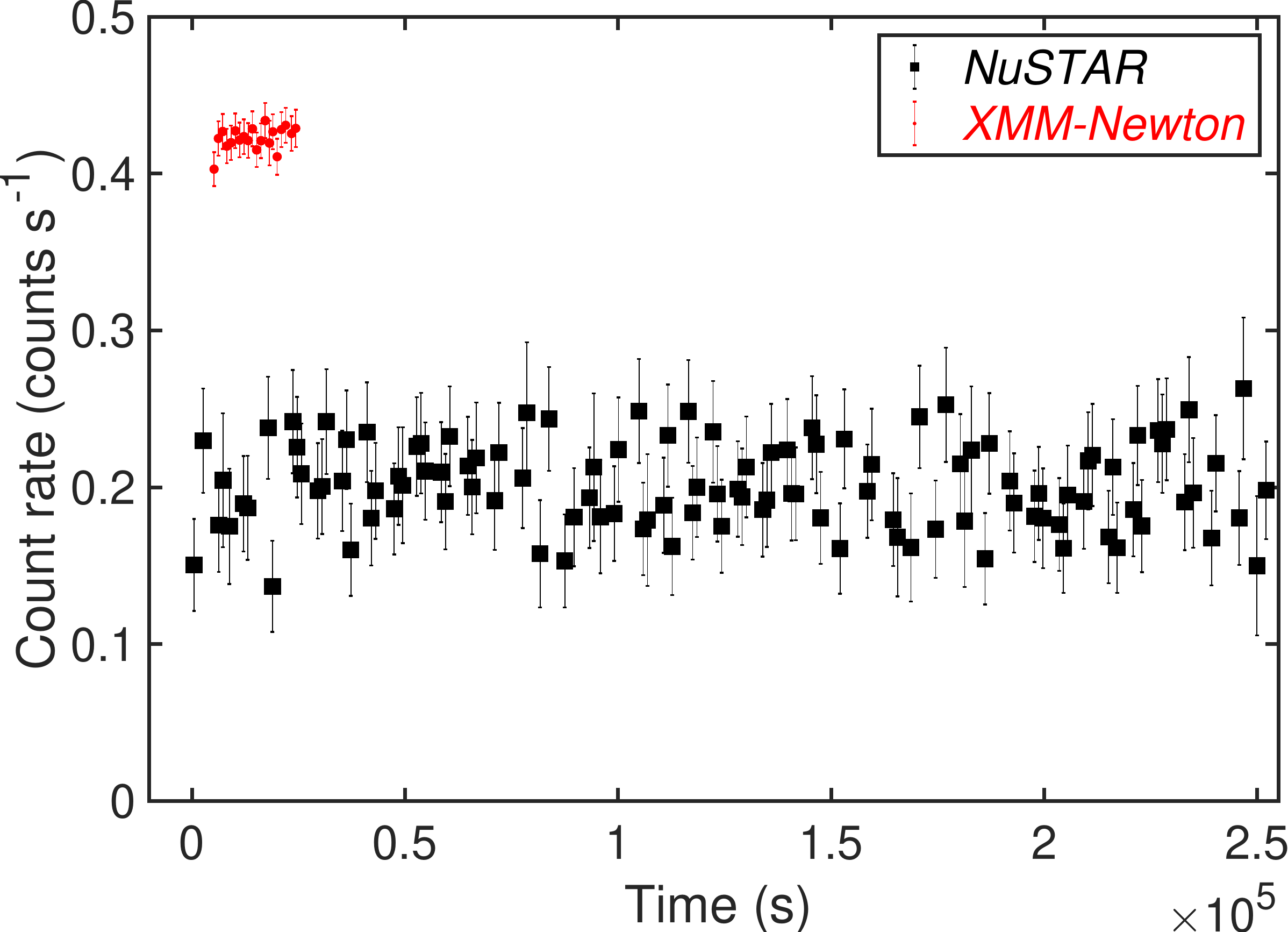}
\caption{{\sl Left panel.} \srco\ \nustar/FPMA and \xmm/PN
  background-corrected light curves binned at 1 ks and 0.2 ks, in the
  energy range 3-60~keV and 0.5-10~keV, respectively. {\sl Right
    panel.} Same for the case of \srct, although both \nustar\ and
  \xmm\ light curves are binned to 1 ks. For clarity, the \xmm\ rate
  has been scaled down by a factor 8 and 7 for \srco\ and \srct,
  respectively. In both panels, time 0 represents the start time of
  the \nustar\ observation.}
\label{lcs}
\end{center}
\end{figure*}

\section{Observations and data reduction}
\label{obs}

The {\it Nuclear Spectroscopic Telescope Array} (\nustar, \citealt{
  harrison13ApJ:NuSTAR}) consists of two identical modules FPMA and
FPMB operating in the energy range 3-79~keV. \nustar\ observed \srco\
on 2016 October 25 for a total, live-time exposure of 104~ks. It
observed \srct\ on 2016 December 06 for 118~ks
(Table~\ref{logObs}). We processed the data using the \nustar\ Data 
Analysis Software, \texttt{nustardas} version v1.8.0 and the
  calibration files CALDB version number 20180419. We reduced the
data using the \texttt{nuproducts} task (which allows for spectral
extraction and generation of ancillary and response files) and HEASOFT
version 6.22.1. We used the flag \texttt{saamode=optimized} to correct
for enhanced background activity visible at the edges of the good time
intervals immediately before entering SAA. We extracted source events
around the source position using a circular region with 60\arcsec
radius, which maximized the S/N ratio. Background events are
extracted from an annulus around the source position with inner and
outer radii of 120\arcsec\ and 200\arcsec, respectively.

%-----------
%Table  2
%-----------
\begin{table*}[ht!]
\caption{\srco\ best-fit spectral parameters}
\label{specParam}
\newcommand\T{\rule{0pt}{2.6ex}}
\newcommand\B{\rule[-1.2ex]{0pt}{0pt}}
\begin{center}
\resizebox{0.90\textwidth}{!}{
\hspace*{-1.0cm}
\begin{tabular}{l c c c c c}
\hline
\srco\ & Cutoff PL & \texttt{pexrav} & \texttt{pexmon} & \texttt{compTT} (slab) & \texttt{compTT} (sphere)\\
\hline
$N_{\rm H}$ ($10^{22}$ cm$^{-2}$) & $0.042\pm0.005$ & $0.045\pm0.005$ & $0.047\pm0.006$ & $0.045\pm0.005$ & $0.044\pm0.005$\\
$kT$ (keV) & -- & -- & -- & $38_{-11}^{+22}$ & $32_{-8}^{+19}$ \\
$\tau$       & -- &  -- & -- &  $1.2\pm0.6$ & $3\pm1$ \\
$\Gamma$ &  $1.79\pm0.01$ & $1.80\pm0.01$ & $1.80\pm0.02$ & -- & -- \\
$E_{\rm cutoff}$~(keV) & $107_{-18}^{+27}$ & $111_{-23}^{+38}$ & $104_{-22}^{+39}$&  -- & -- \\
$R$              & --      &     $<0.06$ & $<0.09$ & -- & -- \\
$F_{\rm 0.5-10~keV}$ ($10^{-11}$ erg s$^{-1}$ cm$^{-2}$) & $1.17\pm0.01$ & $1.16\pm0.01$ & $1.16\pm0.02$ &  $1.17\pm0.01$ & $1.20\pm0.01$\\
$F_{\rm 10-60~keV}$ ($10^{-11}$ erg s$^{-1}$ cm$^{-2}$) & $0.97_{-0.02}^{+0.03}$ & $0.96_{-0.02}^{+0.01}$ & $0.97\pm0.02$& $0.99\pm0.02$ & $1.0\pm0.02$\\
$L_{\rm 0.5-60~keV}$ ($10^{41}$ erg s$^{-1}$) & $5.1\pm0.1$ & $5.0\pm0.1$ & $5.1\pm0.1$ & $5.1\pm0.1$  & $5.2\pm0.1$ \\
\hline
$\chi^2$/d.o.f. \B & 1217/1221 & 1218/1220 & 1217/1220 & 1218/1221 & 1218/1221 \\
\hline
\hline
\end{tabular}}
\end{center}
\end{table*}
%-----------
%Table  2
%-----------

\xmm\ observed both sources simultaneously with \nustar\ for a total,
cleaned EPIC-pn exposure of about 4 and 17~ks for \srco\ and \srct,
respectively (Table~\ref{logObs}). During both observations, the EPIC
cameras \citep{struder01aa} are operated in Full Frame mode, using
the thin filter. The PN and MOS data are selected using event
patterns 0--4 and 0--12, respectively, during only good X-ray events
(``FLAG$=$0''). We inspected all observations for intervals of high
background, e.g., due to solar flares, and excluded those where the
background level was above 5\% of the source flux. We extracted source
events for the two observations from a circle with center obtained by
running the task  {\sl eregionanalyse} on the cleaned event files.
This task calculates the optimum centroid of the count distribution
within a given source region. We set the source extraction radius to
60\arcsec. Background events are extracted from a source-free annulus
centered at the source with inner and outer radii of 120\arcsec and
200\arcsec, respectively. We generated response matrix files using the
SAS task {\sl rmfgen}, while ancillary response files are generated
using the SAS task {\sl arfgen}.

We identified two X-ray point sources within 1.5\arcmin\ of \srco\ with
\xmm. The first, \srco\ X-1, is reported in \citet{ptak04apj:ngc3998}
as a likely background AGN on the basis of its X-ray, UV, and optical
fluxes. The second, \srco\ X-2, is an unidentified source. Both
sources have very low fluxes, $F_{\rm
  0.5-10~keV}\sim10^{-14}$~erg~cm$^{-2}$~s$^{-1}$. \srco\ X-1 is
  not detected above 2~keV, while \srco\ X-2 is very weakly detected,
  with a 2-10 keV flux $\lesssim1\%$ that of the central LLAGN
  (Figure~\ref{skyIm}, left panel). We also detect two weak X-ray
  point sources in the vicinity of \srct. Only one is detected
  $>2$~keV, with a flux that is $\sim$1\% compared to the 2-10~keV
  flux of the nucleus (Figure~\ref{skyIm}, right panel). We conclude
  that neither LLAGN as observed with \nustar\ is contaminated by
  extra-nuclear point sources.

The spectral analysis of the \nustar\ and \xmm\ data was 
performed using Xspec version 12.9.1p \citep{arnaud96conf}. 
The photo-electric cross-sections of \citet{verner96ApJ:crossSect} and
the abundances of \citet{wilms00ApJ} are used throughout to account
for absorption by neutral gas. We bin the spectra to have a S/N ratio
of 6 in each spectral bin, and used the $\chi^2$ statistic in Xspec
for model parameter estimation and error calculation. For all spectral
fits, we added a multiplicative constant normalization between FPMA
and FPMB, frozen to 1 for the former and allowed to vary for the
latter to account for any calibration uncertainties between the two
instruments. We applied the same strategy to account for any calibration
uncertainties between the different EPIC instruments. We find these
calibration uncertainties to be within 2\%. We also allowed for
a constant normalization factor between \nustar\ and \xmm\
instruments, and found that such an uncertainty is around $\sim10$\%.
Finally, all quoted errors throughout the manuscript are at the
$1\sigma$ level, unless otherwise noted.

\begin{figure*}[th!]
\begin{center}
\includegraphics[angle=0,width=0.49\textwidth]{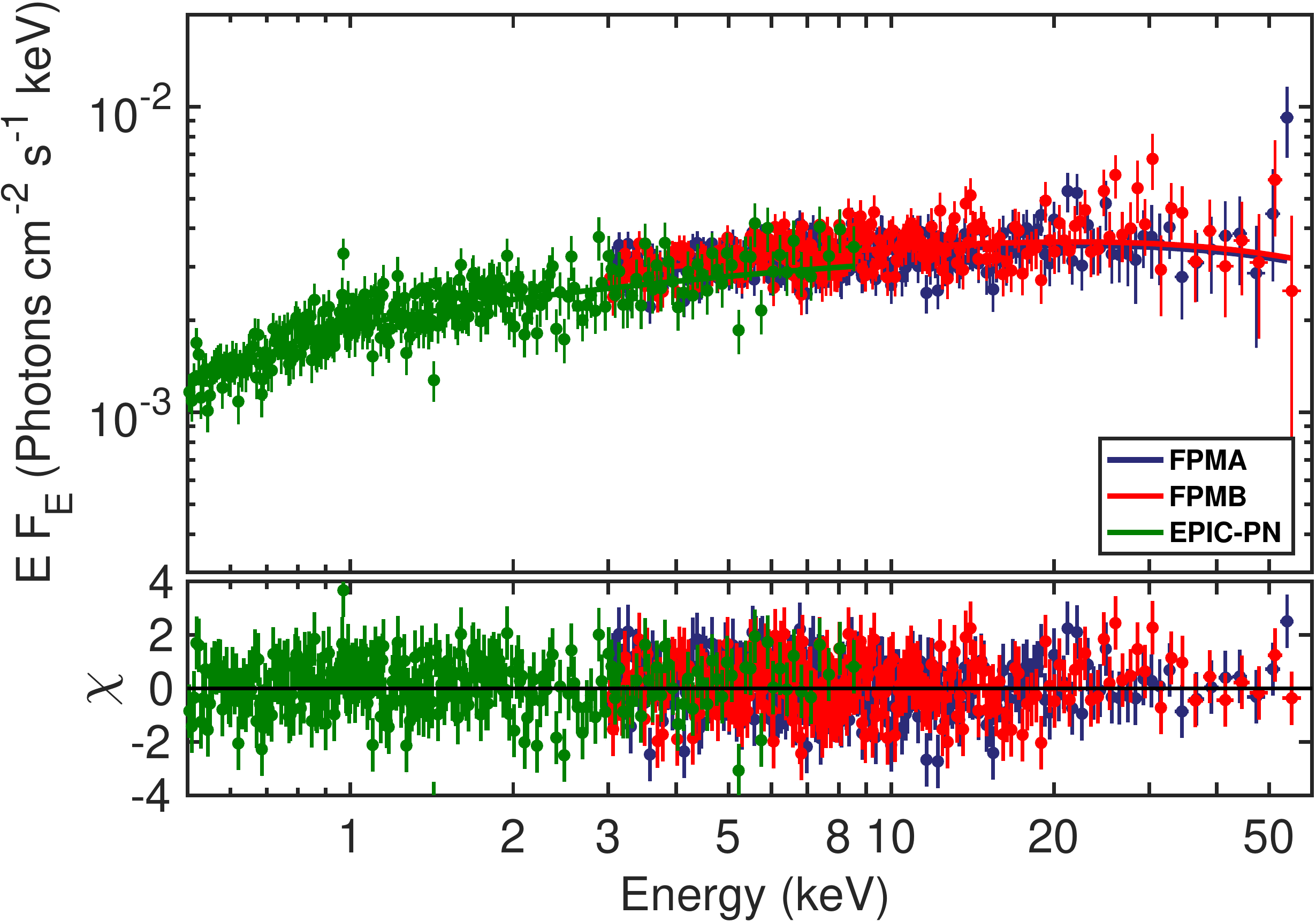}
\includegraphics[angle=0,width=0.49\textwidth]{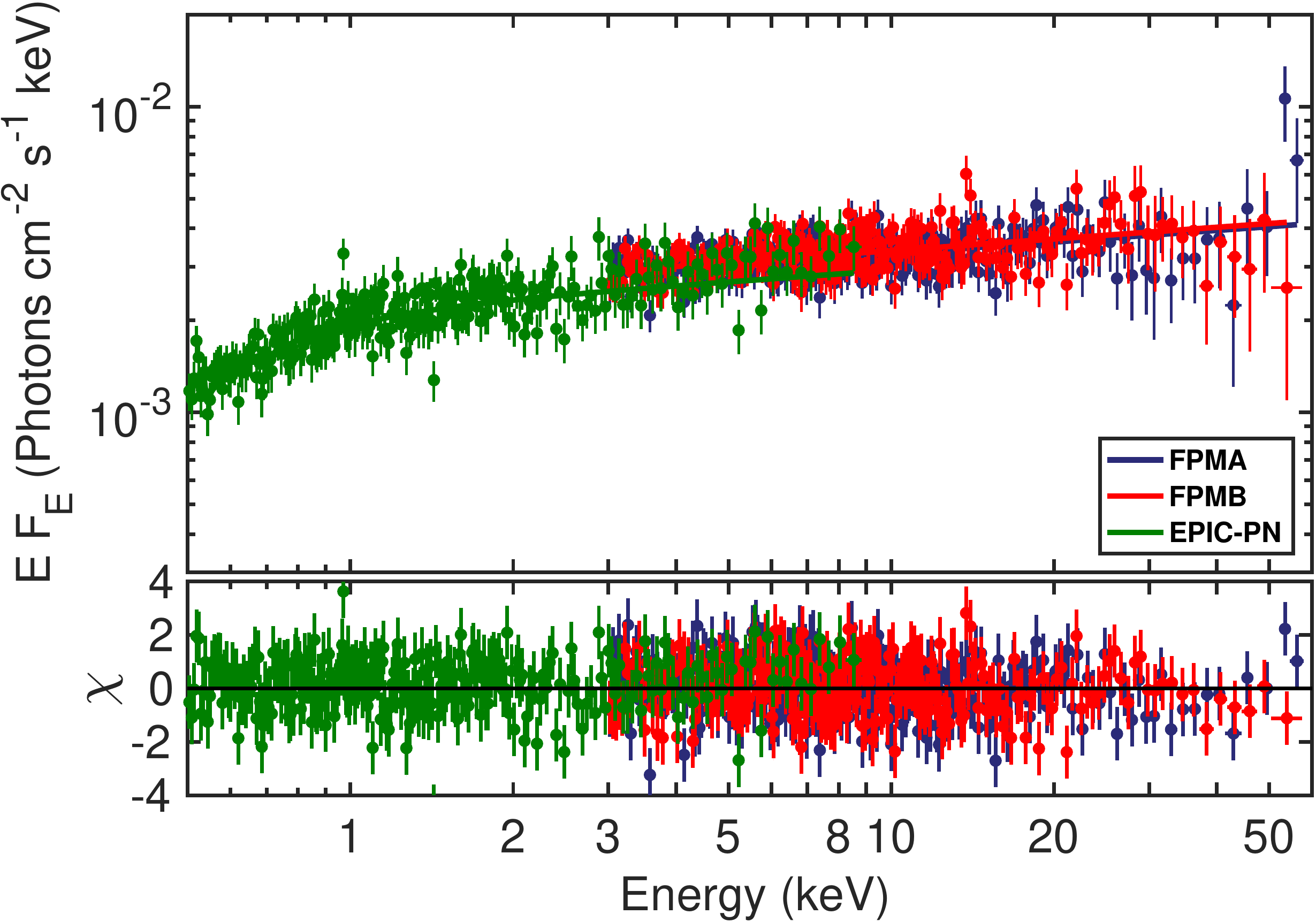}
\caption{{\sl Upper-left panel.} Best-fit cutoff PL model (solid line)
  to the \nustar+\xmm\ spectra of \srco\ (filled circles, only EPIC-pn
  plotted for clarity) shown in $E F_{\rm E}$ space. {\sl Upper-right
    panel.} Best-fit PL model (solid line) to the \nustar+\xmm\
  spectra of \srco\ (filled circles, only EPIC-pn plotted for clarity)
  shown in $E F_{\rm E}$ space. {\sl Lower panels.} Residuals in terms
  of the standard deviation $\sigma$. See text for more details.}
\label{specFit3998}
\end{center}
\end{figure*}

\begin{figure}[th!]
\begin{center}
\includegraphics[angle=0,width=0.49\textwidth]{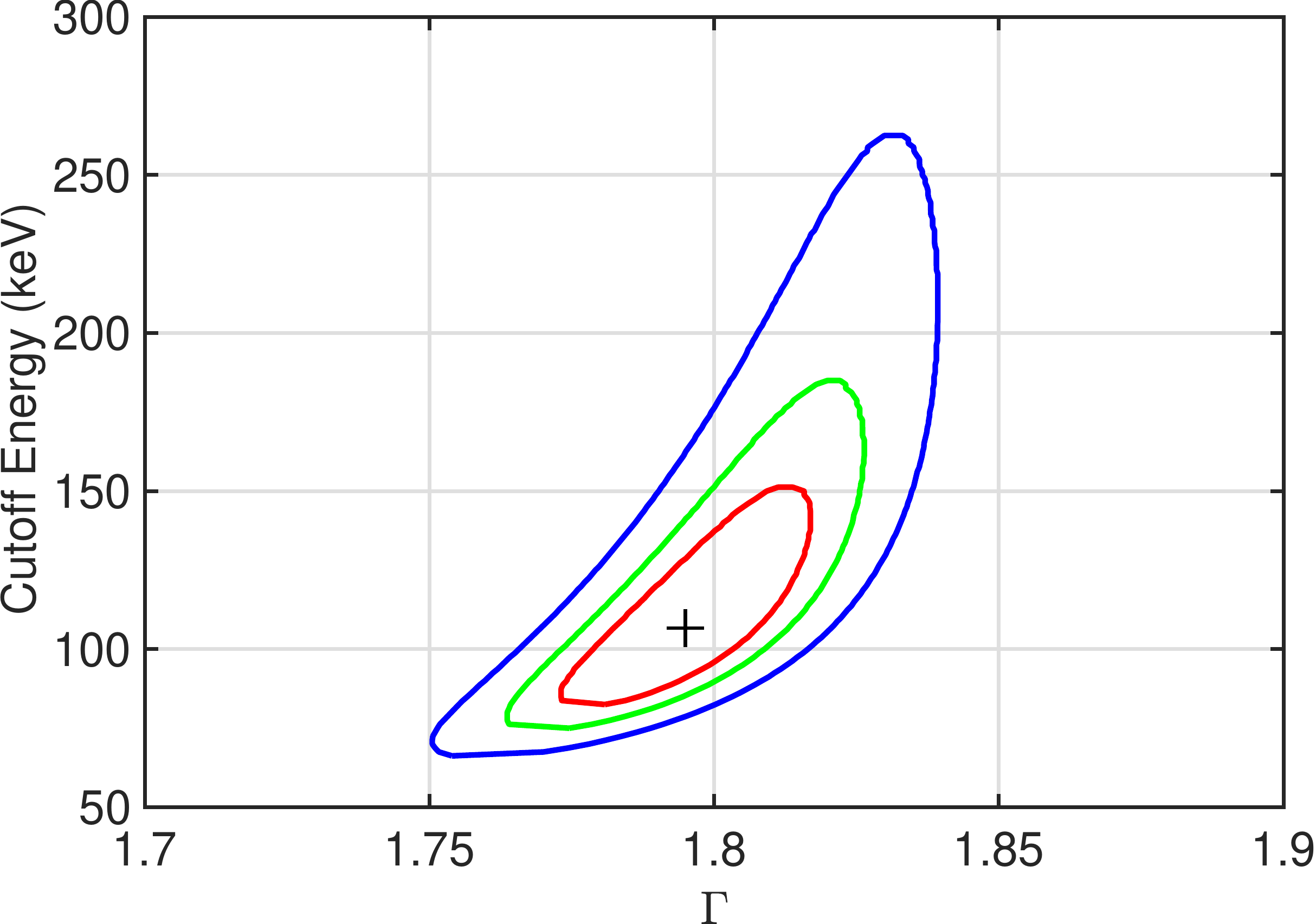}
\caption{\srco\ contour plot of the cutoff energy and PL index
  $\Gamma$; red, green, and blue show the 1, 2, and 3 $\sigma$
  contours, respectively.}
\label{contEcutGam3998}
\end{center}
\end{figure}

\section{Results}
\label{res}

\subsection{Timing analysis}
\label{timeana}

Figure~\ref{lcs} shows the background corrected \nustar\ and \xmm\
light curves, in the energy range 3-60~keV and 0.5-10~keV,
respectively, for both \srco\ and \srct. All light curves are binned
at 1~ks resolution, except for the \xmm\ observation of \srco. It is
binned at 200~s resolution to accomodate the many short good time
intervals as a result of filtering out time intervals of high
background flaring activity. Nonetheless, we find no obvious
variability in either source over the full length of the \xmm\ and
\nustar\ observations; the latter spanning $\gtrsim2.5$~days in
length. Using \xmm, we derive a $3\sigma$ upper limit of
  $\sim$10\% on flux variability on time-scales of 1~ks. On longer
  time-scales of 5 and 10 ks, we derive $3\sigma$ upper limits of
  $\sim$30\% and $\sim$20\% on flux variability using \nustar. For a
more rigorous look at the variability of both sources, we built the
power spectral density (PSD) using the Lomb-Scargle Periodogram and
\nustar\ light curves binned at 60 seconds. The PSD of the two sources
consist of only white noise with no significant red noise component in
the frequency range $\sim10^{-5}-0.017$~Hz.

\begin{figure*}[t]
\begin{center}
\includegraphics[angle=0,width=0.49\textwidth]{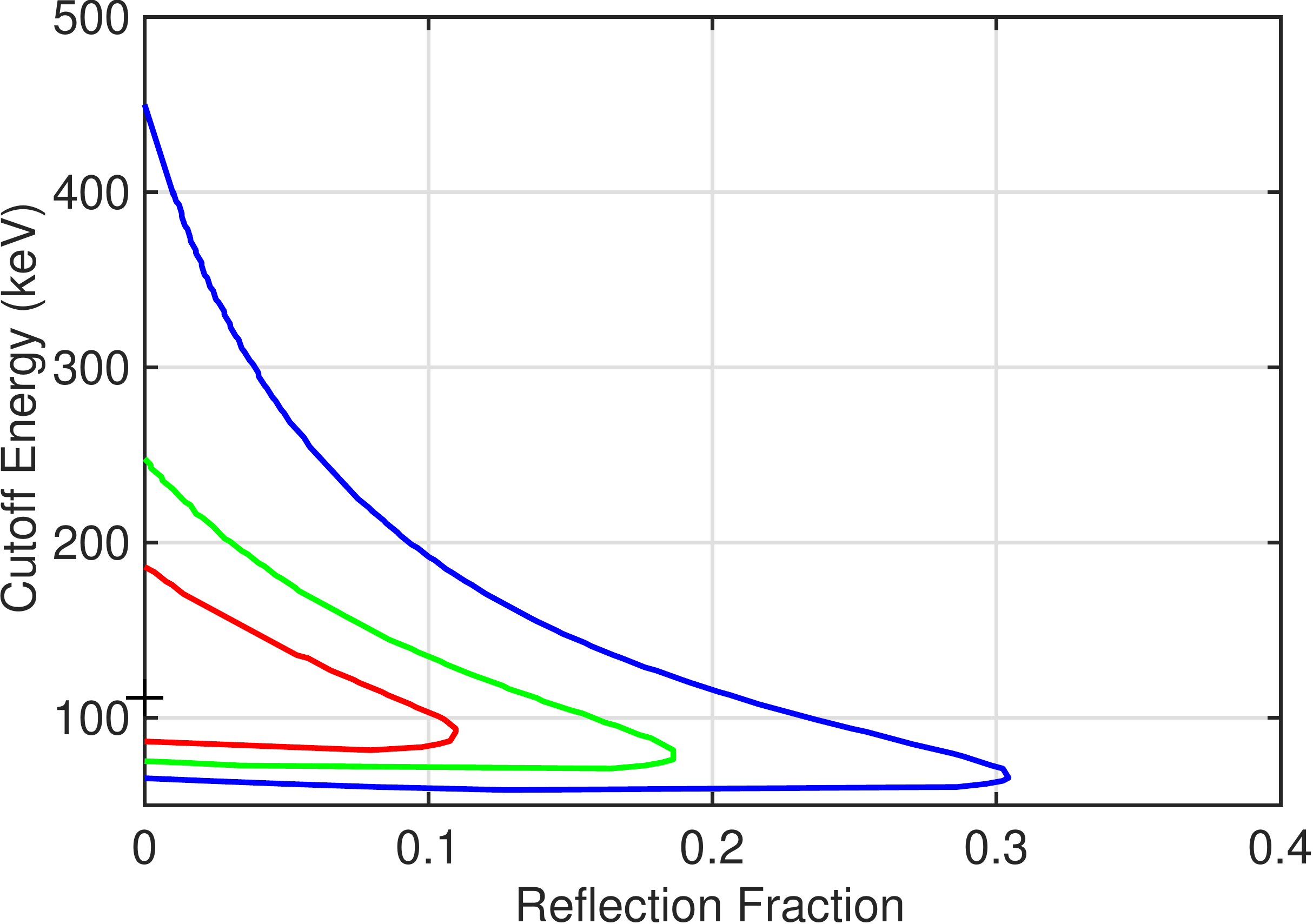}
\includegraphics[angle=0,width=0.49\textwidth]{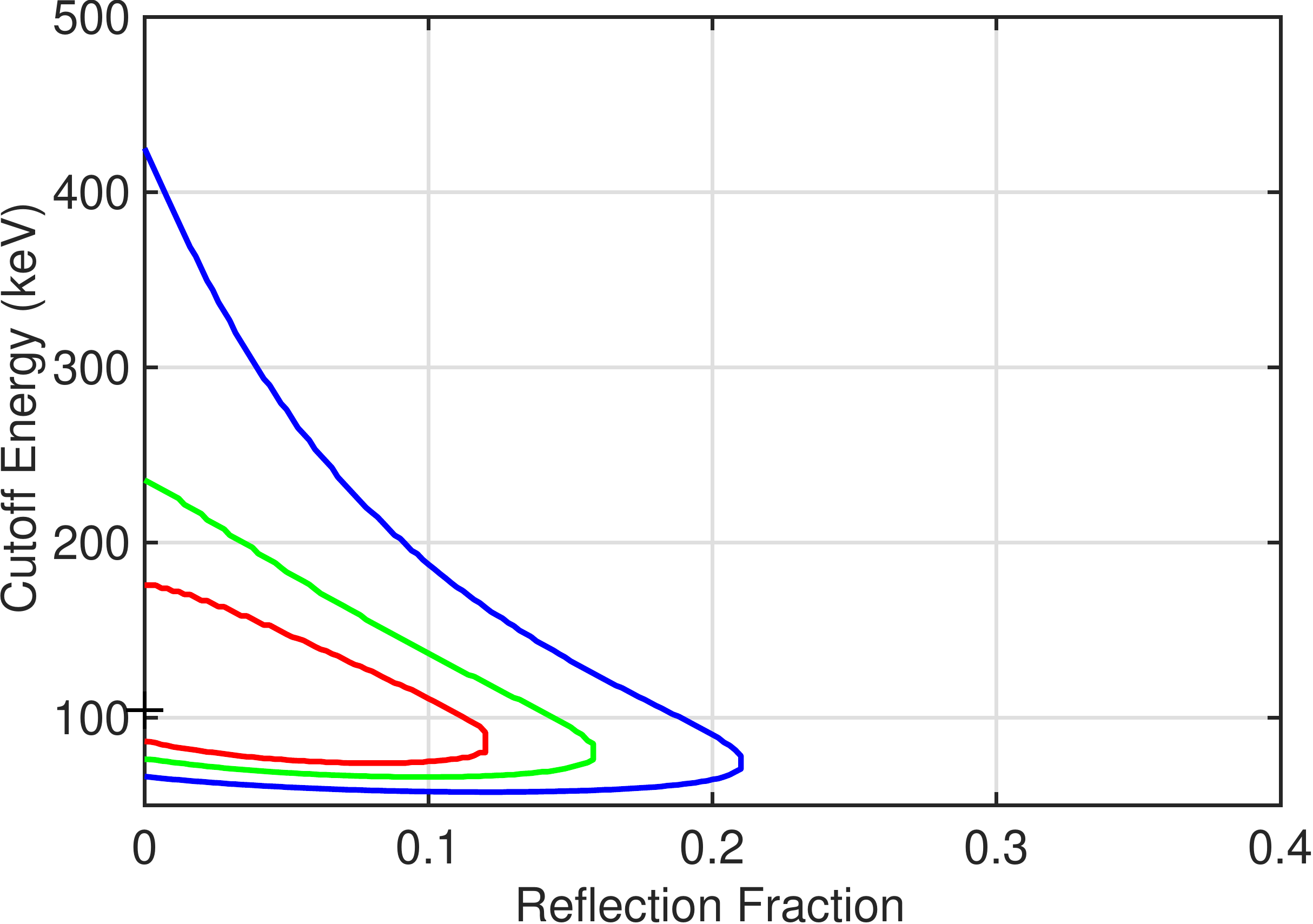}
\caption{Contour plot of the high energy cutoff and reflection
  fraction from optically thick material, $R$. {\sl Left panel} shows
  the results using the \texttt{pexrav} model, while the {\sl right
    panel} shows the effect of using the \texttt{pexmon} model, which
  self-consistently includes reflection from atomic species. Red, green,
  and blue lines show the 1, 2, and 3 $\sigma$ contours, respectively.}
\label{conRefEcut}
\end{center}
\end{figure*}

\begin{figure}[t]
\begin{center}
\includegraphics[angle=0,width=0.49\textwidth]{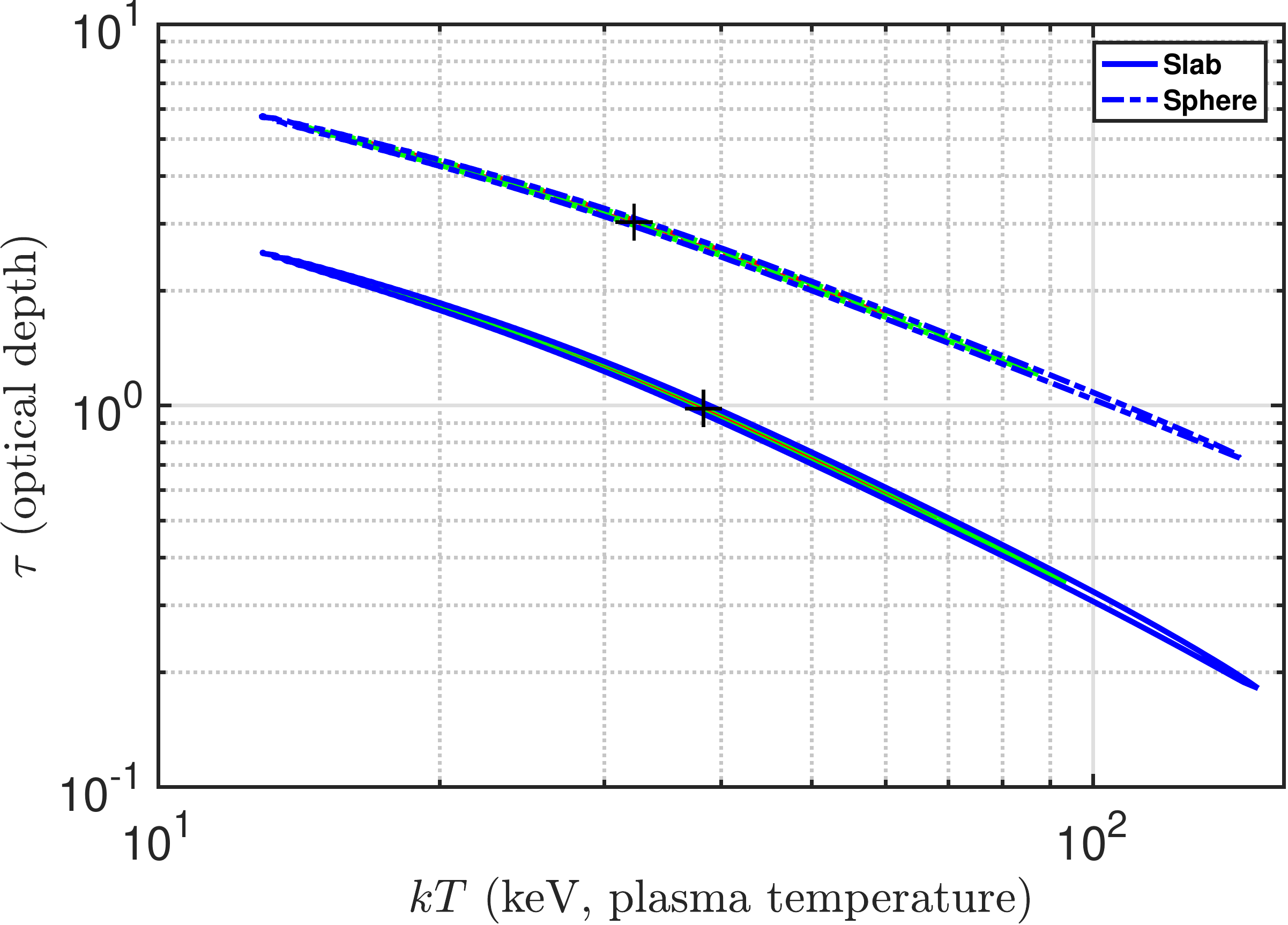}
\caption{Contour plot of the plasma temperature and optical depth for
  the \texttt{compTT} model fit to the \srco\ data, assuming a slab
  geometry (solid lines) and spherical geometry (dashed lines); red,
  green, and blue show the 1, 2, and 3 $\sigma$ contours,
  respectively.}
\label{compcon}
\end{center}
\end{figure}

\subsection{Spectral analysis}
\label{specana}

\subsubsection{\srco}

Before we delved into the spectral analysis of the \nustar+\xmm\
observations, we first checked the highest energy bin in which \srco\
is detected with \nustar. We calculated the number of exposure and 
background-corrected counts from the source location at energies
$>10$~keV, in energy bins of 10~keV. We detect the source in the
energy bin 50-60~keV with a significance of $3.4\sigma$;
$58\pm17$~counts. Above 60~keV, the source becomes indistinguishable
from background. Hence, all the spectral analysis for \srco\ was done
in the 3 to 60~keV range for \nustar\ and 0.5-10~keV for \xmm.

We started our spectral analysis with a simple absorbed (using
\texttt{tbabs} model in Xspec) power-law (PL)
fit to the \nustar+\xmm\ spectra simultaneously. This simple model is
the preferred one to the previous \xmm-only data \citep[e.g., ][]{
  ptak04apj:ngc3998}. We find a statistically acceptable fit with a
$\chi^2$ of 1244 for 1222 degrees of freedom (d.o.f.). We find a small
intrinsic hydrogen column density of about $4\times10^{20}$~cm$^{-2}$,
and a hard photon index $\Gamma=1.86\pm0.01$. Nevertheless, structured
residuals are visible mainly at high energies
(Figure~\ref{specFit3998}). Hence, we replaced the PL with a 
high-energy exponentially-cuttoff one; the \texttt{cutoffpl} model in
Xspec. The reduced $\chi^2$ for this fit is 1217 for 1221 d.o.f. This
is an improvement of $\Delta\chi^2=27$ for 1 additional free
parameter; the cutoff energy. An F-test comparison between the two
models indicates that the probability for the improvement due to the
\texttt{cutoffpl} fit (compared to the one with a single absorbed PL)
to occur by chance is $2.4\times10^{-7}$. We conclude that the cut-off
energy is required by the data and that the \texttt{cutoffpl} model is
the best fit model to the \nustar+\xmm\ spectra of \srco.

The cutoff energy is well constrained by our data $E_{\rm
  cut}=107_{-18}^{+27}$~keV, and the photon index is slightly harder,
$\Gamma=1.79\pm0.01$, compared to the PL model. We find a 0.5-10~keV
flux of about $(1.17\pm0.01)\times10^{-11}$~erg~s$^{-1}$~cm$^{-2}$ and
a slightly lower one,
$0.97_{-0.02}^{+0.03}\times10^{-11}$~erg~s$^{-1}$~cm$^{-2}$, in the
hard 10-60~keV band. The 0.5-10~keV flux that we measure is
consistent, within errors, with the one measured 15 years earlier with
\xmm\ \citep{ptak04apj:ngc3998,younes11AA:liner1sXray}. The best-fit
spectral parameters are summarized in Table~\ref{specParam}, while
the data and best-fit cutoff PL model are shown in
Figure~\ref{specFit3998}. We also show in Figure~\ref{contEcutGam3998} the
1, 2, and 3$\sigma$ contours of the cutoff energy $E_{\rm cut}$ and
the photon index $\Gamma$.

Although we detect no sign of reflection off of dense material,
  e.g., from an accretion disk or a dense molecular torus, we fit the
\nustar+\xmm\ spectra with a reflection model to assess the reflection
fraction limit that we can derive with our data. We used the
\texttt{pexrav} model implemented in Xspec which utilizes as continuum
an exponentially cutoff PL \citep{magdziarz95MNRAS}. The model is also
dependent on the inclination angle of the source, which we fixed to
45~degrees, and on elemental abundances which we assumed to be solar
(we verified that changing these parameters within a valid range does
not affect our results). We find a good fit to the data with this
model with $\chi^2=1218$ for 1220 d.o.f. The result is consistent with
no reflection and we derive a $1\sigma$ upper limit on the reflection
fraction of a fiducial reflection component of 0.06. The continuum fit
parameters are fully consistent with that of the cutoffpl model
fit. We also fit the data using the \texttt{pexmon} model instead
  of pexrav, which has the advantage of self-consistently including
  reflection due to atomic species such as the Fe~K$\alpha$,
  Fe~K$\beta$, and Ni~K$\alpha$ \citep{nandra07MNRAS:pexmon}.  We find
  a statistically equivalent fit with $\chi^2=1217$ for 1220
  d.o.f. The only noticeable difference between the two models is the
  constraint on the reflection fraction, which is shown as contour
  plots in Figure~\ref{conRefEcut}. Due to the non-detection of any Fe
  lines in the spectrum of \srco, the allowed parameter space for the
  reflection fraction in the case of \texttt{pexmon} is more
  stringent. Hence, we consider as conservative the upper-limit on $R$
  as derived with \texttt{pexrav}. The constraint on the cutoff energy
  is similar in the two cases. The best-fit parameters of both models
  are listed in Table~\ref{specParam}.

Finally, we fit the \srco\ spectrum with physically motivated emission
models. Assuming that the accretion geometry is consistent with a hot
flow, the emission process is expected to be either thermal
bremsstrahlung and/or Comptonization of soft photons by the hot plasma
in the flow \citep[see][and references therein]{yuan14ARAA}. Hence, we
fit the spectrum with the thermal bremsstrahlung model
\texttt{zbremss} in Xspec. This model does not give a good fit to the
data with a $\chi^2$ of 2898 for 1222 d.o.f. Fitting the data with a
model consisting of 2 thermal bremsstrahlung model gives a $\chi^2$ of
1294 for 1219 d.o.f.; considerably worse than the cutoff PL model, and
resulting in strong residuals at hard X-rays. We next fit the spectrum
with the Comptonization model \texttt{compTT} in Xspec
\citep{tita94apj}. We assumed that the seed photon temperature is
10~eV. Assuming either a spherical or slab geometry for the Compton
cloud, we get a good fit to the data with a $\chi^2$ of 1218 for 1222
d.o.f., equivalent in goodness to the phenomenological cutoff PL
model. In the spherical geometry case, we get a plasma temperature
$kT=32_{-8}^{+19}$~keV and an optical depth $\tau=3\pm1$, while the
slab geometry results in a plasma temperature $kT=38_{-11}^{+22}$~keV
and an optical depth $\tau=1.2\pm0.6$. The contour plots of the
optical depth and temperature are shown in Figure~\ref{compcon} for
the spherical (dashed lines) and the slab (solid lines) geometries,
respectively. Regardless of the geometry assumed, we find a similar
plasma temperature in the range 15 to 150~keV. The optical depth, on
the other hand, is a factor of 2 to 3 smaller for the slab geometry
compared to the spherical case. We discuss these results in
section~\ref{discuss}. 

\subsubsection{\srct}

Similar to our previous analysis, we first determined the highest
energies at which \srct\ is detected with \nustar. Although the
background was slightly larger in this case, we detect \srct\ up to
60~keV at the $\sim3\sigma$ level, with a background and exposure
corrected number of counts of $51\pm18$.

We fit the \srct\ \nustar+\xmm\ spectra simultaneously, starting
with a simple absorbed PL. The fit is statistically poor, with a
$\chi^2$ of 2312 for 1840 d.o.f. Residuals due to this fit were clear
at low energies, and in the form of emission lines around the Fe-line
complex. Hence, we decided to ignore data in the energy range
5.5-8.0~keV in order to first establish the best-fit continuum model.
We note that the intrinsic absorption to \srct\ was consistent with 0
with a $3\sigma$ upper limit of $0.004\times10^{22}$~cm$^{-2}$,
therefore, in the following we removed the contribution from any
intrinsic absorber. Most early-type galaxies emit an extended hot
diffuse X-ray component usually fit to an emission model from an
optically thin plasma \citep{fabbiano89aa:difgaz}. Hence, we fit the
broad-band spectrum of \srct\ with a combination of a PL and the
Xspec \texttt{mekal} component \citep{mewe85AAS,kaastra93:lines}
  to model the thermal emission from hot diffuse gas in the soft
  band. The fit is statistically good with a $\chi^2$ of 1759 for
1675 d.o.f, and no strong residuals are present at low energies. We
find a plasma temperature $kT=0.63\pm0.02$ and a PL photon index
$\Gamma=1.822\pm0.006$. Replacing the PL model with a cutoff PL, we
find an equally good fit with a $\chi^2$ of 1756 for 1674
d.o.f. Hence, the data does not statistically require a high energy
cutoff. We place a $3\sigma$ upper-limit on any fiducial cutoff
$E_{\rm cut}>152$~keV. For consistency with \srco, in the following,
we utilize the cutoff PL model to establish best-fit parameters for
the Fe line complex. None of the fit parameters are affected if we use
a simple PL model.

\begin{figure}[]
\begin{center}
\includegraphics[angle=0,width=0.49\textwidth]{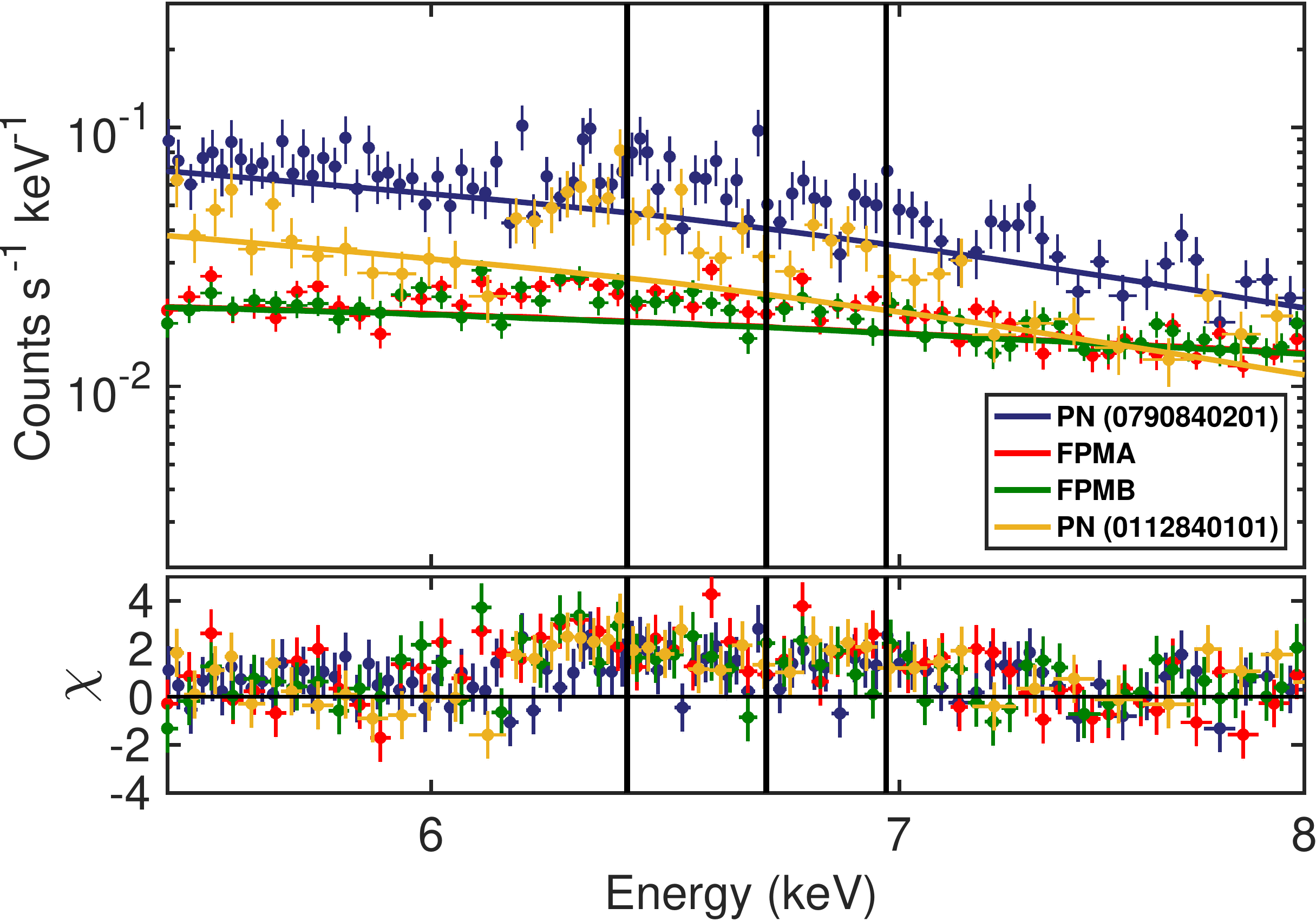}
\caption{{\sl Upper panel.} Count spectrum of our current
  \nustar+\xmm\ observation (red, green, and blue dots, for FPMA,
  FPMB, and pn) and the previous \xmm-only observation (pn, yellow
  dots), zoomed-in at the Fe complex. The solid lines are the best fit
  continuum to all data simultaneously excluding the 5.5-8.0~keV
  energy range. The vertical black lines are plotted at the expected
  energies of the Fe~K (6.4~keV), Fe~XXV (6.7~keV), and Fe~XXVI
  (6.97~keV) lines. {\sl Lower panel.} Residuals in terms of the
  standard deviation $\sigma$. Notice that the residuals deviate from
  the best fit model in the full 6.0-7.0~energy range, indicating an
  Fe complex consisting of multiple ionization species. See text for
  more details.}
\label{FeComplex}
\end{center}
\end{figure}

\begin{figure*}[t!]
\begin{center}
\includegraphics[angle=0,width=0.49\textwidth]{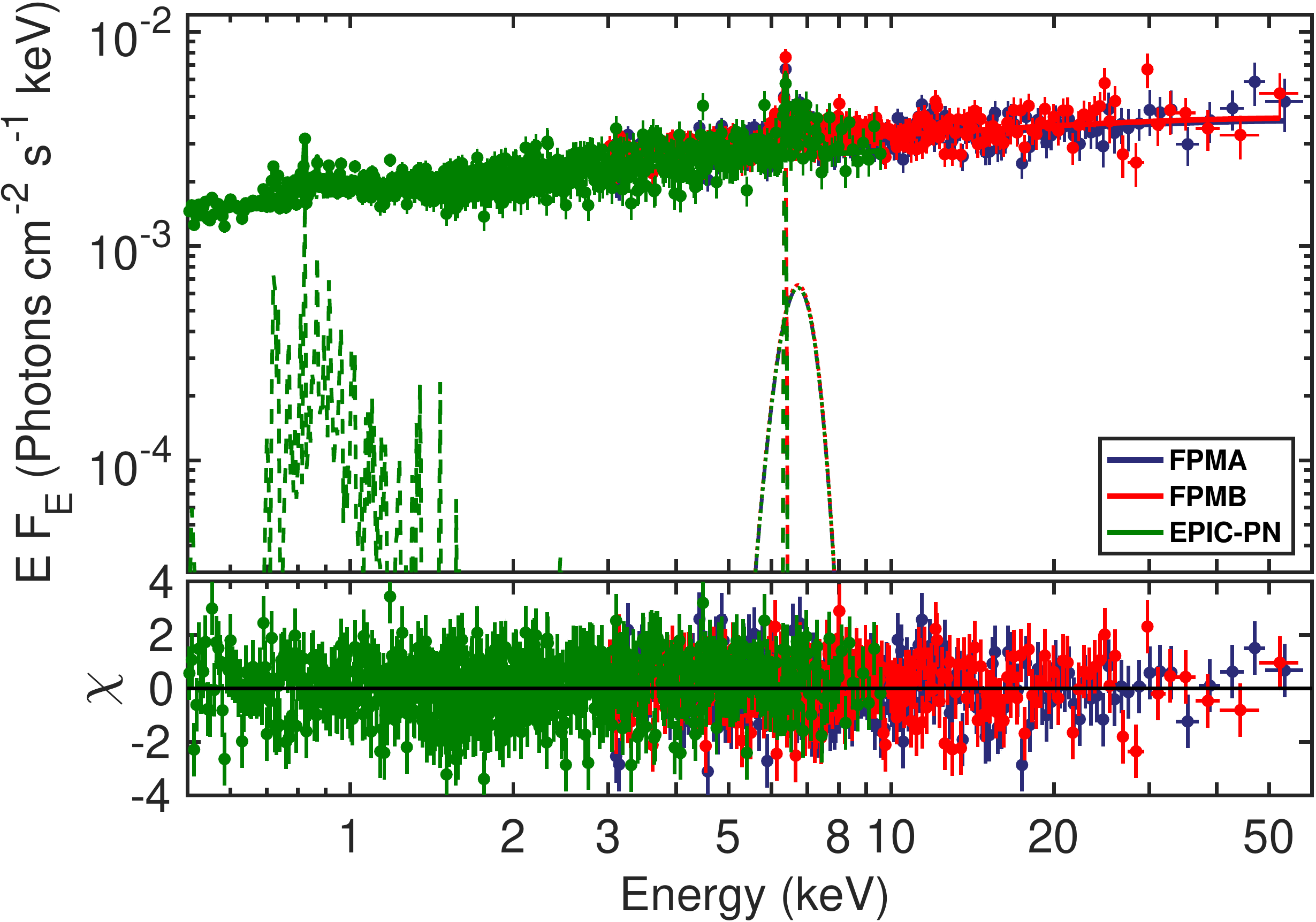}
\includegraphics[angle=0,width=0.49\textwidth]{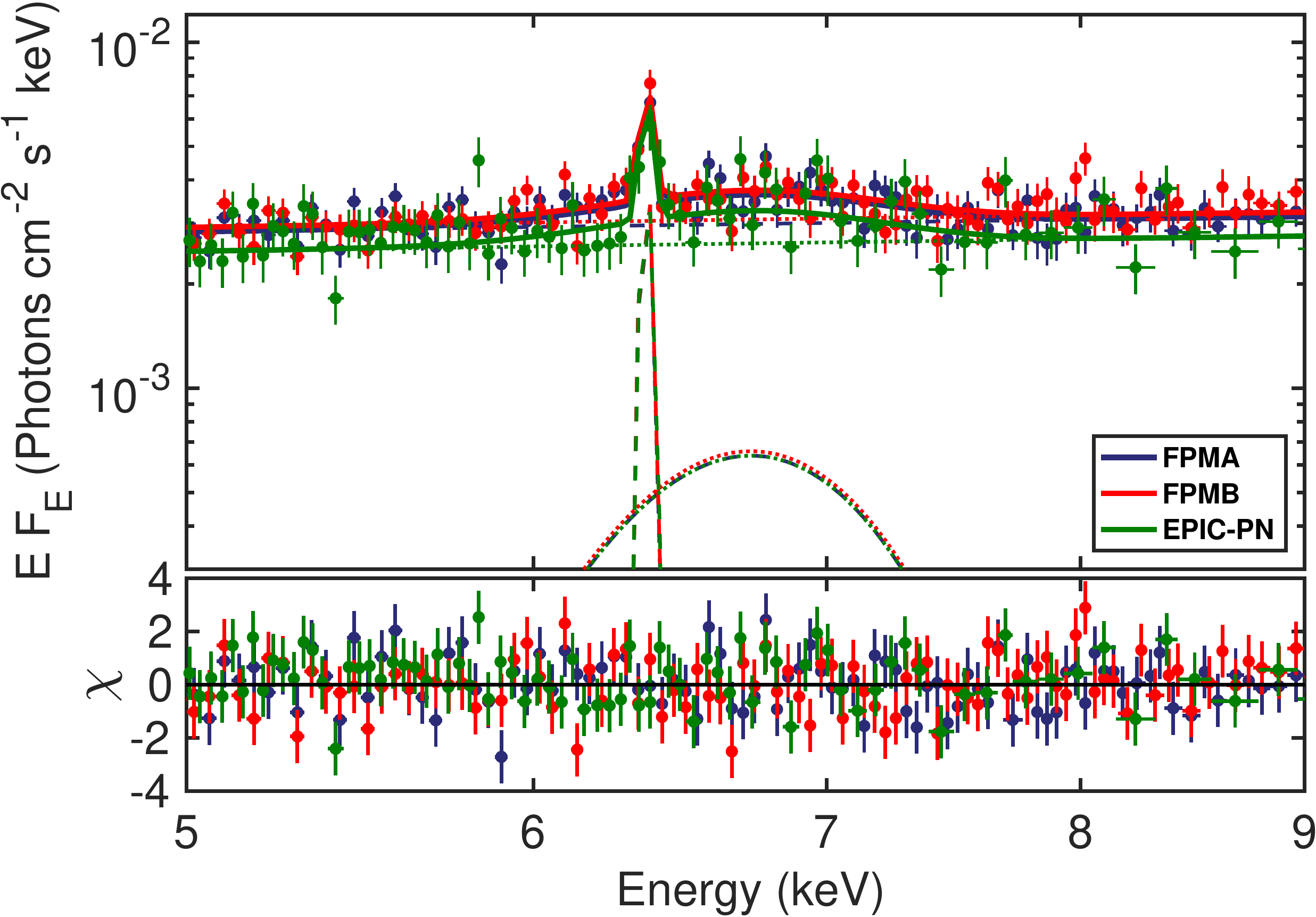}
\caption{{\sl Upper panels.} Best fit model to the \nustar+\xmm\
  spectra of \srct\ (only EPIC-pn shown for clarity). The model
  consists of a cutoff PL, emission from a hot thermal plasma, and two
  Gaussian lines. The different components are shown as dashed lines
  while the sum of all is shown as a solid line. The {\sl left panel}
  shows the broad-band spectrum while a zoom-in at the Fe complex is
  shown in the {\sl right panel}. {\sl Lower panels.} Residuals in
  terms of the standard deviation $\sigma$. See text for more
  details.}
\label{specFit4579}
\end{center}
\end{figure*}

The residuals that we observe in our data around the Fe complex
cover the full $\sim6$ to $\sim7$~keV energy range
(Figure~\ref{FeComplex}). Previously, these residuals have been fit
with either a narrow emission line at $\sim6.4$~keV and a broad
component at slightly higher energies \citep{dewangan04ApJ}, or a
combination of 3 narrow lines at energies fixed at the Fe~K and the
highly ionized Fe species; Fe~XXV and Fe~XXVI \citep{
  terashima98ApJ:4579}. Hence, we firstly added to our continuum model
two Gaussian line profiles. These two components eliminate any
outstanding residuals in the energy range 6 to 7~keV, and we find a
$\chi^2$ of about 1955 for 1880 d.o.f. for the full 0.5-60~keV fit. We
indeed find that the two lines consist of a narrow component at
$6.4\pm0.1$~keV with an unresolved width (fixed to its best fit value
of 0.018~keV), and a broader component with energy centered at
$6.6\pm0.1$~keV, consistent with emission from highly ionized Fe~XXV.
Secondly, we fit the Fe complex with three Gaussian components with
energies fixed at 6.4, 6.7 and 6.97~keV. Without fixing the widths of
the lines there is no unique solution and there is a tendency for the
Fe XXV or Fe XXVI to acquire large widths ($\sim$3.2~keV) and fit part
of the continuum, hence, to test this hypothesis, we also fixed the
widths of these lines to 0, i.e., assuming narrow components. We find
an equally good fit with a $\chi^2$ of 1970 for 1882 d.o.f. 

%-----------
%Table  2
%-----------
\begin{table*}[t!]
\caption{\srct\ best-fit continuum parameters.}
\label{specParam4579}
\newcommand\T{\rule{0pt}{2.6ex}}
\newcommand\B{\rule[-1.2ex]{0pt}{0pt}}
\begin{center}
\resizebox{0.99\textwidth}{!}{
\hspace*{-1.0cm}
\begin{tabular}{l c c c c c}
\hline
\srct\ \T\B & \texttt{mekal+cutoffpl} & \texttt{mekal+pexrav} & \texttt{mekal+pexmon} & \texttt{mekal+compTT} (slab) & \texttt{mekal+compTT} (sphere)\\
\hline
$kT$ (keV, \texttt{compTT}) & -- & -- & -- & $>166$ & $>179$ \\
$\tau$ & -- & -- & -- & $0.05_{-0.02}^{+0.10}$ & $0.31_{-0.16}^{+0.41}$ \\
$kT$ (keV, \texttt{mekal}) & $0.63\pm0.02$ & $0.63\pm0.02$ & $0.62\pm0.02$ & $0.62\pm0.02$ & $0.62\pm0.02$ \\
$\Gamma$ &  $1.81\pm0.01$ & $1.80\pm0.01$ & $1.81\pm0.01$ & -- & -- \\
                   &  $1.87\pm0.01^a$ & -- & -- & -- & -- \\
$E_{\rm cutoff}$~(keV) & $>299$ & $409_{-152}^{+583}$  & $414_{-158}^{+146}$ & -- & -- \\
$R$              & -- & $<0.03$  & $<0.02$ & -- & -- \\
$F_{\rm 0.5-10~keV}$ ($10^{-11}$ erg s$^{-1}$ cm$^{-2}$) & $1.08\pm0.01$ & $1.08\pm0.01$  & $1.07\pm0.01$ & $1.07\pm0.01$ & $1.08\pm0.01$\\
                                                                              & $0.652\pm0.005^a$ & --  & -- & -- & --\\
$F_{\rm 10-60~keV}$ ($10^{-11}$ erg s$^{-1}$ cm$^{-2}$) & $1.04\pm0.02$ & $1.03\pm0.02$ & $1.03\pm0.02$ & $1.02\pm0.02$ & $1.03\pm0.02$ \\
$L_{\rm 0.5-60~keV}$ ($10^{41}$ erg s$^{-1}$) & $6.8\pm0.1$ & $6.8\pm0.1$ & $6.7\pm0.1$ & $6.7\pm0.1$ & $6.8\pm0.1$ \\
\hline
$\chi^2$/d.o.f. \B & 1955/1880 & 1953/1880 & 1954/1880 & 1955/1880 & 1955/1880\\
\hline
\hline
\end{tabular}}
\end{center}
\begin{list}{}{}
\item[{\bf Notes.}] $^a$ Spectral properties of the previous \xmm\
  observation of \srct.
\end{list}
\end{table*}
%-----------
%Table  2
%-----------

%-----------
%Table  3
%-----------
\begin{table}[t!]
\caption{Best fit Gaussians to the Fe complex.}
\label{specParamFeLines}
\newcommand\T{\rule{0pt}{2.6ex}}
\newcommand\B{\rule[-1.2ex]{0pt}{0pt}}
\begin{center}
\resizebox{0.49\textwidth}{!}{
\hspace*{-1.0cm}
\begin{tabular}{l c c c c}
\hline
Model & $E$ & $\sigma$ & $N$ & EW \\
          & keV  & keV & ($10^{-6}$ photons cm$^{-2}$ s$^{-1}$) & eV \\
\hline
Two Gaussians & $6.4\pm0.1$ &  0.018(f) & $4\pm1$ & $52_{-19}^{+13}$ \\
                        & $6.6\pm0.1$ & $0.5\pm0.1$ & $19\pm3$ & $256_{-31}^{+36}$ \\
\hline
\hline
Three Gaussians & $6.4$(f) &  0.0(f) & $9\pm1$ & $118\pm14$\\
                           & $6.7$(f) &  0.0(f) & $4\pm1$ & $43\pm11$\\
                           & $6.97$(f) &  0.0(f) & $5\pm1$ & $72_{-14}^{+11}$\\
\hline
\hline
\end{tabular}}
\end{center}
% \begin{list}{}{}
% \item[{\bf Notes.}] $^a$ Line properties when including the previous
%   \xmm\ observation of \srct.
% \end{list}
\end{table}
%-----------
%Table  3
%-----------

\begin{figure*}[t]
\begin{center}
\includegraphics[angle=0,width=0.49\textwidth]{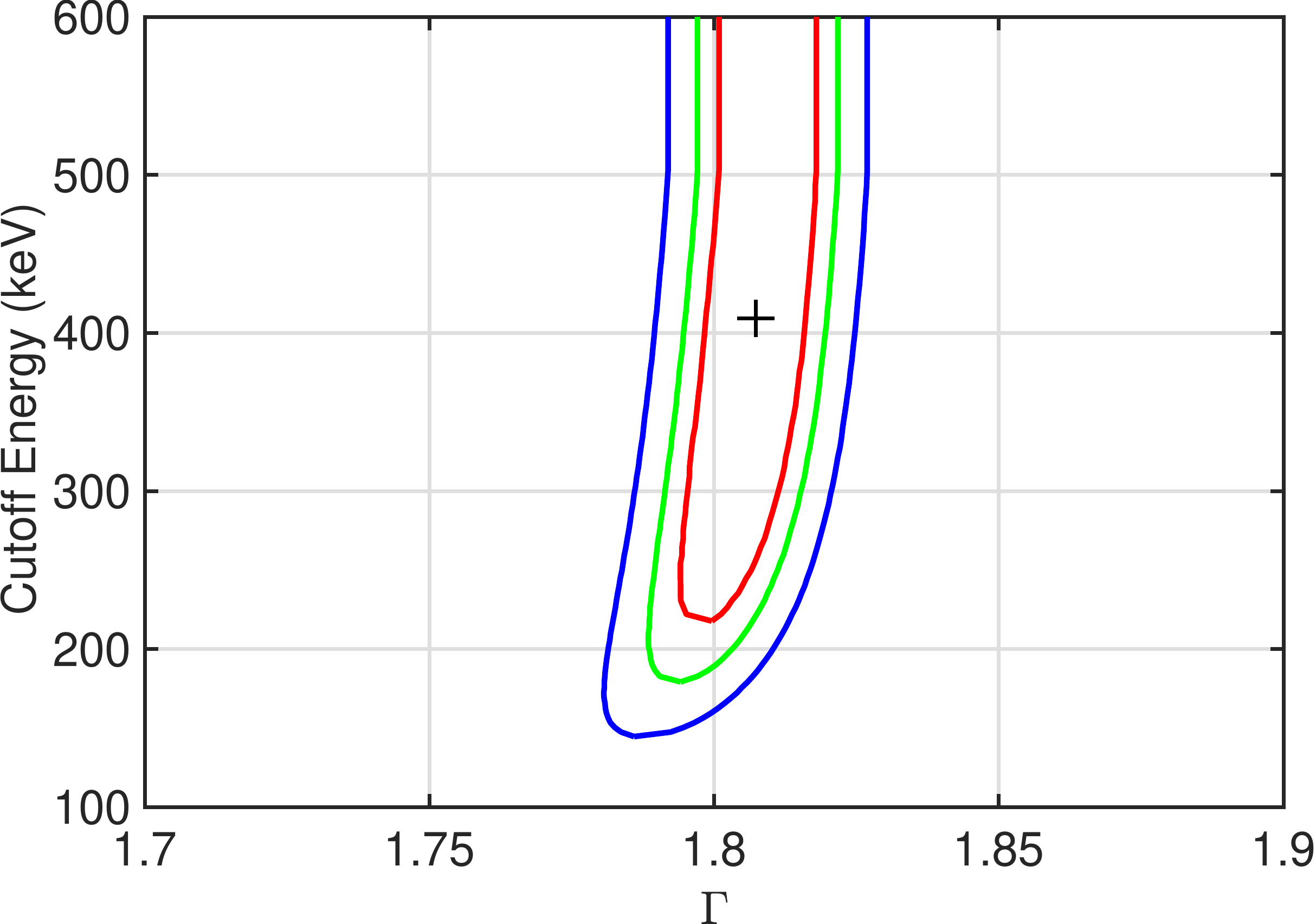}
\includegraphics[angle=0,width=0.49\textwidth]{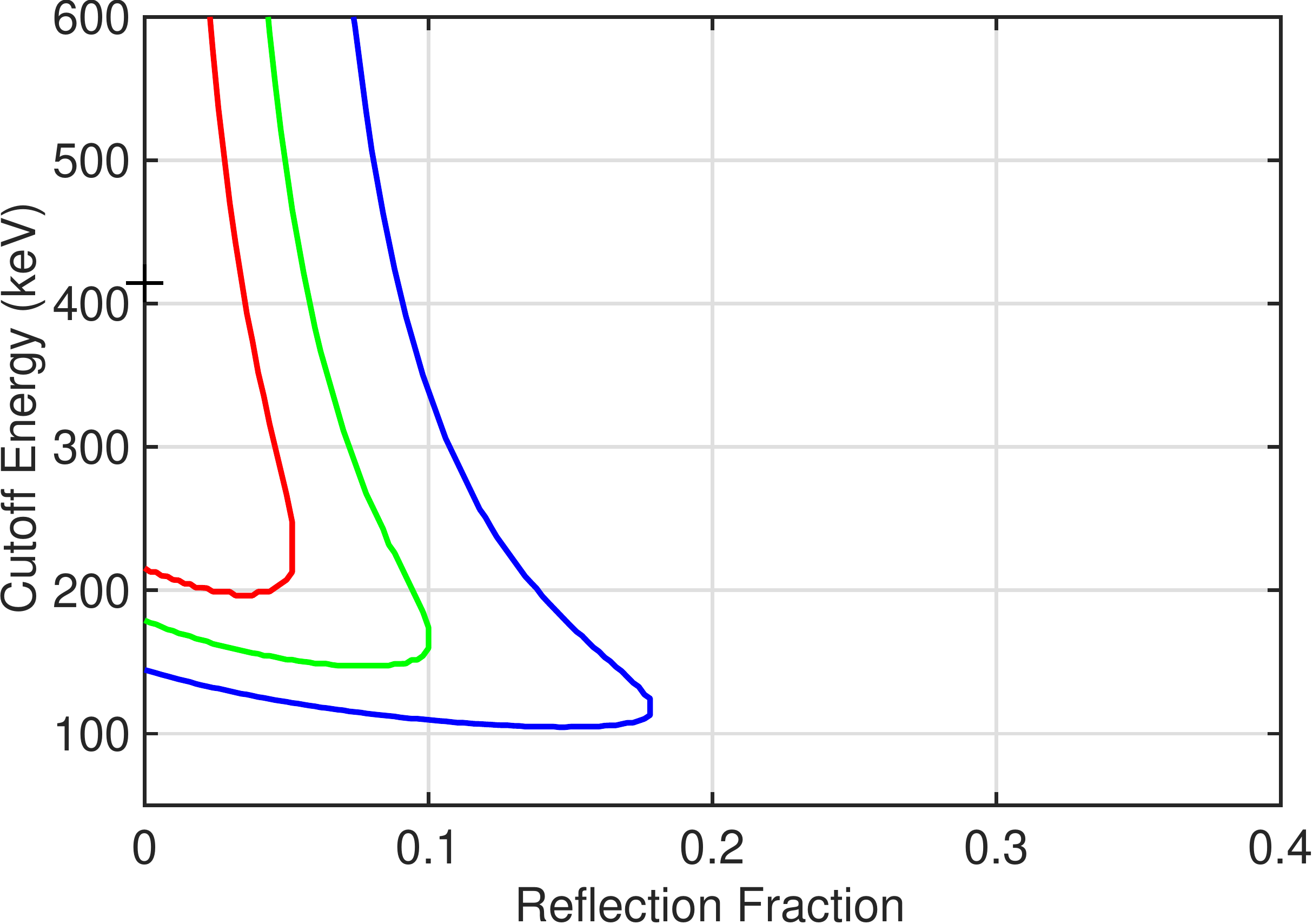}
\caption{\srct\ $E_{\rm cut}-\Gamma$ contour plot ({\sl left panel}),
  and $E_{\rm cut}-R$ contour plot ({\sl right panel}). Red, green,
  and blue show the 1, 2, and 3 $\sigma$ contours, respectively. See
  text for more details.}
\label{conRefEcut4579}
\end{center}
\end{figure*}

Figure~\ref{specFit4579} shows the broad-band best-fit model and a
zoom-in at the Fe line complex fit with 2 Gaussian components (the
residuals due to a fit with 3 narrow Gaussians look fairly similar).
We report the best-fit model parameters of the continuum in
Table~\ref{specParam4579}, while the Fe lines best fit parameters are
reported in Table~\ref{specParamFeLines}. Figure~\ref{conRefEcut4579}
shows the 1, 2, and 3 $\sigma$ contours of $E_{\rm cut}$ and
$\Gamma$.

Similar to \srco, we do not detect any sign of reflection off of a
geometrically thick accretion disk. Nonetheless, in order to establish
an upper-limit on any fiducial reflection component, we replace the
cutoff PL with the \texttt{pexrav} model in fitting the broad-band 
spectrum of \srct. We fixed the inclination angle  of the disk to
45~degrees, and assumed solar abundances. We find a good fit to the 
data with a $\chi^2=1953$ for 1880 d.o.f. We find a $1\sigma$ upper
limit on the reflection $R\lesssim0.03$. The continuum fit parameters
are fully consistent with that of the cutoffpl model fit. The best-fit
parameters are listed in Table~\ref{specParam} and we show in
Figure~\ref{conRefEcut4579} the 1, 2, and 3 $\sigma$ contours of $R$
and $E_{\rm cut}$. These results are discussed in Section~\ref{discuss}.

We also attempted to fit the \srct\ spectrum with the
  \texttt{pexmon} model. We removed the neutral Gaussian component
  from the model, which is self-consistently produced by
  \texttt{pexmon} in Compton-thick material. This fit results in a
  $\chi^2$ of 1975 for 1881 d.o.f. Strong residuals are seen at high
  energies above 15 keV where the model overestimates the flux from
  the source.

Given the possible Compton-thin origin of the neutral Fe line
  that we see in \srct, we also fit the spectrum with the
  \texttt{MYTorus} model \citep{murphy09MNRAS,yaqoob12MNRAS}. This
  model includes the reprocessed continuum and neutral Fe~K$\alpha$,
  Fe~K$\beta$, and Ni~K$\alpha$ emission by a torus-shaped structure
  with an opening angle of 60\degree. The continuum is assumed a PL
  with a high-energy cutoff fixed at 500~keV, and the column densities
  for the \texttt{MYTorus} tables reproducing the scattered continuum
  and emission lines are tied. A {\texttt mekal} component was also
  included to the continuum model. We also included
  contributions from narrow Fe~XXV and Fe~XXVI by fixing the 2
  Gaussian components to their expected energies and their widths to
  0. The fit is good with a $\chi^2$ of 1919 for 1880 d.o.f. and we do
  not observe any unmodeled residuals around the Fe complex. The
  hydrogen column density of the material responsible for the
  reprocessed continuum and the production of the neutral Fe~K line is
  $N_{\rm H}=(7\pm2)\times10^{22}$~cm$^{-2}$ (errors quoted at the
  3$\sigma$ level). We conclude that the line is likely originating
  from a Compton-thin material at large distances from the central
  BH.

\begin{figure}[t]
\begin{center}
\includegraphics[angle=0,width=0.49\textwidth]{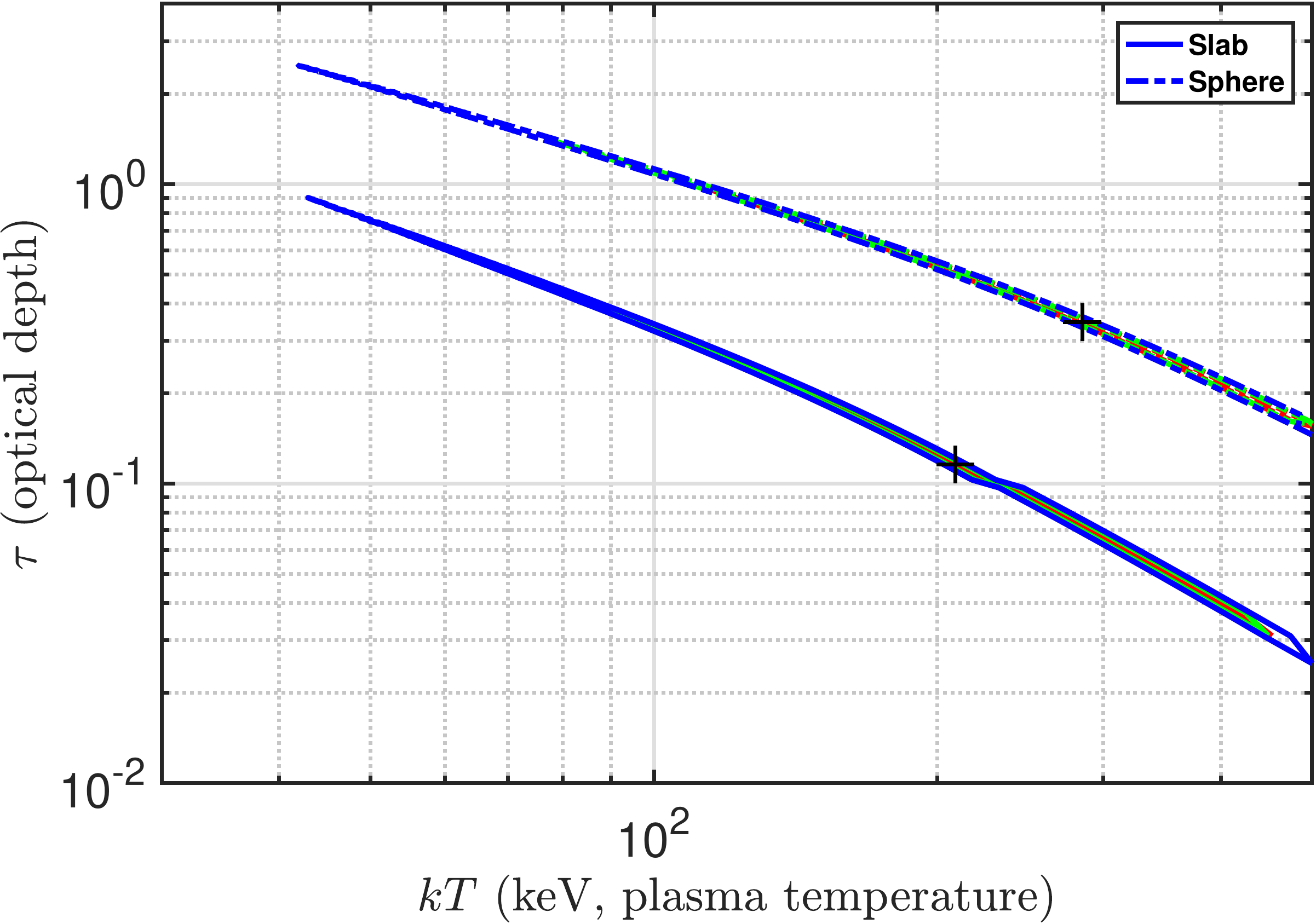}
\caption{Contour plot of the plasma temperature and optical depth for
  the \texttt{compTT} model fit to the \srct\ data, assuming a slab
  geometry (solid lines) and spherical geometry (dashed lines); red,
  green, and blue show the 1, 2, and 3 $\sigma$ contours,
  respectively.}
\label{conTauTmp4579}
\end{center}
\end{figure}

Although we do not detect a cutoff in \srct, we fit the
  broad-band spectrum of the source with a physical Comptonization
  model, \texttt{compTT} in Xspec, to establish the allowed parameter
  space of the optical depth and temperature of the X-ray emitting
  plasma, and to allow comparison with \srco\ and, more broadly,
  luminous AGN. We find a good fit to the spectrum with a $\chi^2$ of
  1955 for 1880 d.o.f., comparable to our best fit with a cutoff
  PL. We find a 1$\sigma$ lower limit on the plasma temperature of
  $\sim170$~keV for either the slab or spherical geometry. We find an
  optical depth $\lesssim1$ in both cases, indicating an optically thin
  plasma. The best-fit parameters are summarized in
  Table~\ref{specParam4579} and we show the contours $\tau$ versus
  $kT$ in Figure~\ref{conTauTmp4579}.

\begin{figure}[t]
\begin{center}
\includegraphics[angle=0,width=0.49\textwidth]{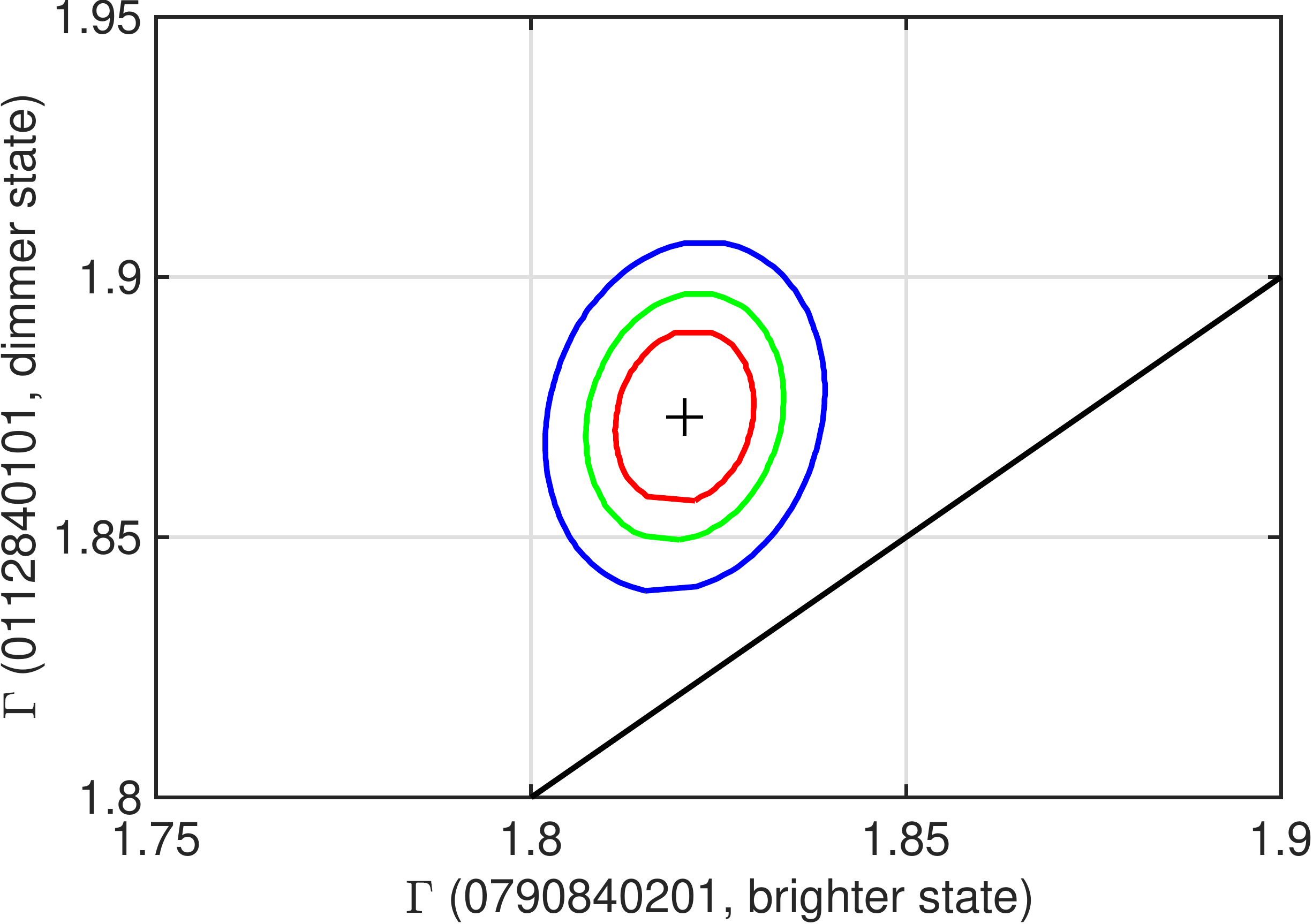}
\caption{Contour plot of the photon indices derived from our
  simultaneous fit of the current \xmm+\nustar\ and previous \xmm\
  observations of \srct. The solid line denote the 1 to 1 relation.
  The brighter flux state of our current observation corresponds to
  the harder X-ray spectrum.}
\label{phot1phot2}
\end{center}
\end{figure}

The 0.5-10~keV flux that we derive in our observation is 60\%
larger than the one derived in the previous \xmm\ observation of
\srct\ \citep{dewangan04ApJ}. Hence, to search for any spectral
variability in concordance to the brighter flux, we reanalyzed the
historic \xmm\ observation taken on 2004 June 30 (obs. ID 0112840101)
as described in Section~\ref{obs}. We fit the spectra of the current
and previous observations simultaneously with a continuum consistent
with the one discussed above, i.e., a hot thermal plasma model and a
cutoff PL (we fixed the cutoff energy for the previous \xmm\
observation to 1 MeV). We linked the thermal spectral components
between the 2 observations since no variability is expected in the hot
diffuse gas over years time-scales. We let the cutoff PL index free to
vary. Firstly, we exclude the Fe complex between 5.5 and 8 keV. The
residuals in this energy range compared to the best fit continuum are
shown in Figure~\ref{FeComplex}. We fit the Fe complex with either 2
or 3 Gaussian components, in the same manner we conducted the spectral
analysis of our current observation. We let the Gaussian parameters
free to vary between the 2 observations. Within statistical
uncertainties we do not detect any variability in the Fe lines. Hence,
we link the Gaussian parameters between the previous and the current
observation. On the other hand, we detect a significant variability in
the PL photon index, with the previous, dimmer observation possessing
a softer spectrum. Figure~\ref{phot1phot2} shows the contour plot of
the indices from the two observations. We summarize in
Table~\ref{specParam4579} the continuum spectral parameters of this
fit.

\section{Discussion}
\label{discuss}

The superior sensitivity of \nustar\ is allowing, for the first time,
a detailed look at the hard X-ray view of LLAGN, e.g.,  NGC~7213 and
M~81 \citep{ursini15MNRAS:7213,young18MNRAS:m81}. In this paper, we
add to this growing sample by presenting the analysis of the
broad-band, 0.5-60~keV, X-ray emission from two LLAGN, \srco\ and
\srct, observed simultaneously with \nustar\ and \xmm. The \nustar\
live-time exposure of the two sources are about 100 and 120~ks,
respectively, taken over 2.5 and 3~days. This allowed us to look for
variability at days time-scale, rarely available for LLAGN. We do not detect any
strong variability in either source. Their power spectra in the 3-60~keV
band  is consistent with white noise in the frequency range
$\sim10^{-5}-0.017$~Hz. This confirms that indeed most LLAGN show
little to no variability on days time-scales \citep{binder09apj:ngc3226,
  younes11AA:liner1sXray,young18MNRAS:m81}, as opposed to the strong,
short time-scale variability seen in bright Seyfert galaxies
\citep[e.g.,][]{gonzalez12AA:var}.

The broad-band 0.5-60~keV continua of \srco\ and \srct\ are best fit
with a cutoff PL (discussed in detail below) and a PL plus emission
from hot thermally-emitting plasma, respectively. Very small intrinsic
absorption is required for \srco, while for \srct\ the absorption
column density is consistent with the Galactic one. This is in line 
with the picture of LLAGN being devote of absorbing material, such as
a broad-line region and/or a torus in their central engine \citep{
  ho08aa:review}. Moreover, \citet{she2018ApJllagn} showed a positive
correlation between the absorption column density and the Eddington
ratio, $\lambda_{\rm Edd}$, in a sample of LLAGN, in striking contrast
to luminous AGN \citep{ricci17Natur}, but consistent with the
expectation of absorption caused by outflowing material from a hot
accretion flow \citep{yuan15ApJ}. Our two sources nicely follow the
correlation, especially \srco, where we indeed measure a very small
intrinsic absorption corresponding to its $\lambda_{\rm
  Edd}\approx10^{-5}$.

Neither source shows any hint of Compton reflection hump at hard
X-rays, a ubiquitous feature in luminous AGN. The $3\sigma$
upper-limits that we derive on their reflection fractions, 0.30 and
0.18 for \srco\ and \srct, respectively, are strongly constraining and
place the two sources within the same $R$-Eddington ratio parameter
space as other LLAGN, e.g., M81 \citep[$R<0.19$,][]{young18MNRAS:m81},
NGC~7231 \citep[$R<0.20$,][]{ursini15MNRAS:7213}, and the low
Eddington ratio AGN NGC~2110 \citep[$\lambda_{\rm Edd}\lesssim10^{-2},
R<0.25$,][]{marinucci15MNRAS}. These very small reflection fractions
are at odds with the much higher values measured for luminous AGN
\citep[e.g.,][]{risaliti13Natur,parker14MNRAS, kara15MNRAS:0707,
  keck15ApJ:4151}. With the addition of \srco\ and \srct, there is now
mounting evidence that the lack of reflection in the spectra of LLAGN,
if not universal, is a common feature of the population. The nature of
the hard X-ray Compton hump in luminous AGN is believed to be
reflection of the primary continuum source from the inner regions of a
dense, optically thick accretion disk extending close to the innermost
stable circular orbit \citep{george91MNRAS}. The absence of this
signature component from the spectra of LLAGN reinforces the idea that
most likely their inner accretion disk is truncated at a certain large
radius from the BH, which is instead filled with an optically thin,
radiatively inefficient hot accretion flow \citep[e.g.,][]{
  esin97ApJ:adaf,narayan97apjl,yuan07ASPC:adaf,yuan14ARAA}. Moreover,
the absence of the Compton hump is consistent with the lack of another
reflection component in LLAGN; the broad Fe~K$\alpha$ line
\citep[e.g.,][]{gonzalezmartin09aa,younes11AA:liner1sXray}. The latter
is also ubiquitous in the X-ray spectra of bright AGN, and have been
shown to correlate temporally with the Compton hump implying similar
physical nature \citep{kara15MNRAS:rev}.

We confirm the existence of the Fe complex in \srct, which was also
detected with \asca\ \citep{terashima98ApJ:4579,terashima00ApJ:4579}
and the previous \xmm\ observation of the source \citep{
  dewangan04ApJ}. The residuals in our present observation indicate a
complex blending of lines, which requires at least 2 Gaussian
components. If the energies and the widths of the Gaussians are left
free to vary, the residuals are best fit with an unresolved, narrow
Fe~K line at 6.4~keV and a moderately broad component consistent with
highly ionized Fe XXV at $\sim$6.7~keV. Nevertheless, the complex can
also be well fit with three, relatively narrow Gaussian components
with energies fixed to 6.4, 6.7, and 6.97~keV, corresponding to Fe~K,
and highly ionized Fe~XXV and Fe~XXVI. 

The narrow neutral component corresponding to Fe~K$\alpha$ is
  unlikely to originate in Compton-thick material. Fitting the \srct\
  spectrum with the reflection model \texttt{pexmon}, which
  self-consistently includes contribution from neutral Fe originating
  in Compton-thick material, does not adequately fit the data. We
  observe strong residuals at high energies above 20 keV where the
  model overestimates the data. This result may imply that the
  formation of the observed Fe~K line in Compton-thick material
  requires a strong reflection component at high energies that is not
  detected in the spectrum. On the other hand, the \texttt{MYTorus}
  model gives an adequate fit to the data, including the narrow Fe~K
  feature. According to this model, the line is most likely produced
  in an optically thin region with a hydrogen column density
  $\sim7\times10^{22}$~cm$^{-2}$ (e.g., the broad-line region;
  \citealt{yaqoob04ApJ}). This is consistent with the presumed
  production site of the line in other LLAGN, namely M~81 and NGC7213
  \citep{young18MNRAS:m81,ursini15MNRAS:7213}. \srct, similar to many
  other LLAGN, show strong signatures of silicate emission features in
  the mid-IR strongly indicative of a dusty environment \citep{
    gallimore10ApJS:IRAGN,mason12AJ:IRLINER}. Such environments are
  also possible sites for forming Fe~K$\alpha$ lines; dust may help
  enhance the Fe~K equivalent width while suppressing strong
  high-energy reflection components due to a decrease in the
  backscattering opacity in the gas
  \citep{draine03ApJ:dustXray}. Although to that end, we note that
  \srco\ does show strong silicate features with no indication of Fe~K
  emission \citep{sturm05ApJ:3998,mason12AJ:IRLINER}.

The highly ionized Fe~XXV and Fe~XXVI lines are most likely the result
of collisional ionization in optically thin gas, e.g., possibly at the
transition layer between a truncated external thin accretion disk and
a hot inner accretion flow. The temperature at the transition layer is
supposedly $\sim10^7-10^9$~K \citep{narayan97apjl,perna00ApJ:adaf},
i.e., the temperatures required to produce highly ionized Fe. These
lines can also be broad with equivalent width as large as a few
hundred eV \citep{xu11ApJ}. Such models have been shown to naturally
explain the Fe~XXV and Fe~XXVI lines seen in M~81
\citep{young07ApJ:m81} and the Fe~XXV line seen in Sgr~A$^*$
\citep{xu06ApJ:sgra,wang13Sci:sgrA}. Another possibile production site
is the outflows from the accretion flow. In this scenario, blueshift
corresponding to the outflow velocity is expected. Unfortunately,
energy resolution and small number statistics renders the distinction
between the above two possibilities difficult.

The current observation of \srct\ shows a factor $\sim$60\% larger
flux compared to the previous \xmm\ observation taken 13 years
earlier. At the same time, the source spectrum shows a significant
hardening (Figure~\ref{phot1phot2}). This is in line with the
harder-when-brighter correlation established for different samples of
LLAGN \citep[e.g.,][]{younes11AA:liner1sXray,she18ApJ:llagn}. Assuming
that the brighter flux is the result of increased accretion rate onto
the SMBH, the spectral hardening can be understood in the context of
radiatively inefficient, optically thin, hot accretion flows. The
increased densities in the flow will lead to more efficient inverse
Compton scattering of primary synchrotron photons from the flow
leading to a harder spectrum \citep[e.g.,][]{qiao13ApJ,yang15MNRAS}.
Within statistical uncertainties, we do not find any variability in
the Fe line complex between the two observations.

The \nustar+\xmm\ data of \srco\ revealed the need for a high
energy cutoff to best fit its broad-band 0.5-60~keV spectrum. We
measure a cutoff energy $E_{\rm cut}=107_{-18}^{+27}$~keV. This
represents the lowest and best constrained cutoff energy ever
measured for a LLAGN. At first glance, this cutoff energy resembles
the ones derived for Seyfert Galaxies and other luminous AGN. However,
for normal AGN, the cutoff energy is found to be inversely
proportional to the Eddington ratio (\citealt{ricci18MNRAS:ecut};
however, note that this relation is not seen when a much smaller
catalog of bright AGN observed with \nustar\ is considered;
\citealt{tortosa18AA}). Sources with Eddington ratio, $\lambda_{\rm
Edd}=L_{\rm bol}/L_{\rm Edd}<0.1$ tend to have a cutoff energy of
about 370~keV, while the ones with $\lambda_{\rm Edd}>0.1$ possess a 
much smaller cutoff energy, $E_{\rm cut}=160$~keV. \srco, hence,
represents an obvious outlier to this  correlation, lying in the
parameter space of low cutoff energy with very small Eddington
ratio.

The negative correlation as observed for normal AGN could be
understood in the context of compact, X-ray emitting corona
\citep{haardt93ApJ:corona}. The energy exchange becomes more efficient
between photons and particles with decreasing size and/or increasing
luminosity of the X-ray emitting corona, which cools down its plasma
content leading to a decrease in temperature. Following the
definition of the compactness parameter $l\propto\lambda_{\rm
  Edd}/R_{\rm X}$ \citep[e.g.,][]{fabian15MNRAS,fabian17MNRAS,
  ricci18MNRAS:ecut}, and assuming a typical luminous AGN corona size
$R_{\rm X}\approx10~R_{\rm G}$~where $R_{\rm G}$ is the gravitational
radius, we estimate $l_{\rm \srco}\approx0.01$. This can be thought of
as a rough upper-limit as $R_{\rm X}$ can indeed be larger for the
LLAGN case. On the other hand, we find electron temperatures
$\theta=kT_{\rm e}/m_{\rm e}c^2$ in the range of $0.03-0.3$ from our
Comptonization spectral fits\footnote{The electron temperatures that
  we derive through the Comptonization model do indeed follow the
  relation $kT_{\rm e}=E_{\rm cut}/2$ for optically thin plasma, where
  $E_{\rm cut}$ is the cutoff energy established through our cutoff PL
  modeling.}. This small compactness parameter along with the small
electron temperature places \srco\ in uncharted territory in the
compactness-temperature plane \citep{fabian15MNRAS}, closer to the
electron-proton coupling line and bremsstrahlung as cooling modes
rather than electron-electron coupling. The latter is the preferred
radiation process for luminous AGN. This could imply that the physical
properties of the X-ray emitting region of \srco, and possibly most
LLAGN, differs markedly from luminous AGN. The very small compactness
parameter and radiation process argues for the expectation of a large,
optically thin hot accretion flow. 

We note that using the lower limits that we derive from our
  Comptonization spectral fits to the \srct\ spectra, we derive an
  electron temperature $\theta\gtrsim0.1$ and a compactness parameter
  $\lesssim 0.1$. These limits fall within the range of
  electron-proton coupling line. However, since only upper-limits can
  be derived in such a case, the spectrum could still be consistent
  with electron-electron coupling similar to what is observed for
  bright AGN. Future soft MeV missions are crucial to constrain the
  high-energy cutoff in LLAGN spectra, and better understand the
  physical properties of the high-energy emitting region in these
  sources.

The high quality, simultaneous \nustar+\xmm\ data allowed us to fit
physically-motivated models to the broad-band spectrum of \srco. A
bremsstrahlung-only model does not give an adequate fit to the
data. In fact, we also tried to fit the spectrum with 2 bremsstrahlung
components, akin emission from a 2 temperature hot accretion flow, and
this too resulted in a statistically poor fit. The Comptonization
model \texttt{compTT}, on the other hand, resulted in a good fit, with
best fit statistics equivalent to the one with the phenomenological
cutoff PL model. Both a spherical geometry for the Compton cloud and a
slab geometry resulted in equally good fits. The contour plots of
these fits, as shown in Figure~\ref{compcon}, indicate a high level of
degeneracy between the electron temperature and the optical depth, a
fact already known for Comptonization models \citep[e.g.,][]{
  brenneman14ApJI,balokovic15ApJ}. Nonetheless, within the $3\sigma$
level, both geometries result in a plasma temperature ranging from
15~keV to 150~keV. The optical depth, however, differs between the two
geometries, and ranges between 0.8 and 6 for the spherical geometry
and 0.2 to 2.5 for the slab geometry. Smaller optical depths for the
slab geometry compared to the spherical case have already been pointed
out (in, e.g., \citealt{luinski10MNRAS}, \citealt{brenneman14ApJI},
\citealt{balokovic15ApJ}), and are the result of integration over
radial distances and scale heights, respectively. These results
support the picture of Comptonization as the dominant emission process
over bremsstrahlung in \srco. Our results are also consistent with
Comptonization emission from optically-thin hot accretion flows
(e.g., ADAF, \citealt{yuan07ASPC:adaf}), which predict high
temperatures of $\sim100$~keV and small optical depths $\tau\lesssim1$
(derived after integrating vertically, hence, to be compared with the
slab geometry results). Finally, this low-energy cutoff detection in
\srco\ limits the validity range of jet emission models. Pure
synchrotron emission from a jet is inherently difficult to produce a
cutoff at hard X-rays \citep{zdziarski04MNRAS}, nevertheless,
Comptonization in the base of a jet has been shown to produce such
curvature in the X-ray spectra of hard state X-ray binaries
\citep[e.g.,][]{ markoff05ApJ}.

\section{Summary}
\label{conc}

In the present work, we report on the broad-band X-ray timing and
spectral analysis of the two LLAGN \srco\ and \srct\ through
simultaneous \nustar+\xmm\ observations. The summary of our main
results are the following:

\begin{itemize}[noitemsep]
\item We do not detect any significant variability in either source
  over the $\sim3$-day length of the \nustar\ observations; both
  sources have power spectra consisting of white noise in the
  frequency range $\sim10^{-5}-0.017$~Hz.
\item The broad-band 0.5-60~keV spectrum of \srco\ is best fit with a
  cutoff power-law; cutoff energy $E_{\rm cut}=107_{-18}^{+27}$~keV.
  This represents the lowest and best constrained high-energy cutoff
  ever measured for a LLAGN. Such relatively low value places \srco\
  as an outlier to the anticorrelation found in luminous AGN between
  $E_{\rm cut}$ and the Eddington ratio.
\item \srco\ spectrum is consistent with a Comptonization model with
  either a sphere or slab geometry with optical depths in the range of
  0.8-6 and 0.2-2.5, respectively, corresponding to plasma
  temperatures between 20 and 150~keV. Its spectrum is inconsistent
  with bremsstrahlung emission.
\item The broad-band 0.5-60~keV spectrum of \srct\ is best fit with a
  combination of a hot thermal plasma model, a power-law, and a blend
  of Gaussians to fit an Fe complex observed between 6 and
  7~keV. These residuals could either be fit with a narrow Fe~K line
  at 6.4~keV and a moderately broad Fe~XXV line, or 3 relatively
  narrow lines, which includes contribution from Fe~XXVI.
\item \srct\ flux is 60\% brighter than previously detected with \xmm,
  accompanied by a hardening in the spectrum.
\item Neither source shows any reflection hump with a $3\sigma$
  reflection fraction upper-limits $R<0.3$ and $R<0.18$ for \srco\
  and \srct, respectively.
\end{itemize}

The very low reflection fractions that we derive for our two sources,
along with the lack of variability and broad Fe~K$\alpha$ lines, argue
for an altered accretion geometry in LLAGN compared to luminous
AGN. This reinforces the picture of a truncated thin disk which is
replaced by a pressure-dominated hot accretion flow. This picture is
also in line with our finding of a harder-when-brighter spectrum for
\srct. Finally, our most interesting result, the relatively low
high-energy cutoff that we measure for \srco\ ($E_{\rm
  cut}=107_{-18}^{+27}$~keV), is inconsistent with the picture of
X-ray emission emanating from a compact corona regulated through
electron-electron coupling, as is widely accepted for luminous
AGN. The combination of very low compactness and 20-150~keV plasma
temperature that we derive is more in line with emission from a large
optically thin volume where electron-proton coupling and
bremsstrahlung act as cooling modes. Such picture is again in favor of
emission from an optically-thin hot accretion flow.

\section*{Acknowledgments}

This work made use of data from the \nustar\ mission, a project led by
the California Institute of Technology, managed by the Jet Propulsion
Laboratory, and funded by the National Aeronautics and Space 
Administration. We thank the \nustar\ Operations, Software and
Calibration teams for support with the execution and analysis of these
observations. This research has made use of the \nustar\ Data Analysis
Software (NuSTARDAS) jointly developed by the ASI Science Data Center
(ASDC, Italy) and the California Institute of Technology (USA). GY
acknowledges support from NASA under \nustar\ Guest Observer cycle-2,
proposal number 15-NUSTAR215-0024. LCH was supported by the National
Key R\&D Program of China (2016YFA0400702) and the National Science
Foundation of China (11473002, 11721303). Y.T. is supported by JSPS
Grants-in-Aid for Scientific Research 15H02070 and 16K05296. FGX and
FY are supported in part by National Key R\&D Program of China (grants
2016YFA0400804 and 2016YFA0400704), the Natural Science Foundation
of China (grants 11873074, 11573051, 11633006, 11650110427,
11661161012, 11303008, and 11473002), and the Key Research Program of
Frontier Sciences of CAS (grants QYZDJSSW-SYS008 and QYZDB-SSW-SYS033
). FGX is also supported by the Youth Innovation Promotion Association
of CAS (id. 2016243) and the Natural Science Foundation of Shanghai
(No. 17ZR1435800). We thank the referee for a concise and thorough
reading of the manuscript that led to an improved version of the article.

\end{document}